\documentclass[aps,prd,twocolumn,superscriptaddress,showpacs]{revtex4}
\usepackage{epsfig,epsf}
\usepackage{amsmath}
\usepackage{amsthm}
\usepackage{amsfonts}
\usepackage{amssymb}
\usepackage{dsfont}
\usepackage{multirow}
\usepackage{appendix}
\usepackage{slashed}
\usepackage[active]{srcltx}
\usepackage{psfrag}
\usepackage{subfigure}
\newcommand{\be}{\begin{equation}}
\newcommand{\ee}{\end{equation}}
\newcommand{\ba}{\begin{eqnarray}}
\newcommand{\ea}{\end{eqnarray}}
\newcommand{\baa}{\begin{eqnarray*}}
\newcommand{\btab}{\begin{tabular}}
\newcommand{\etab}{\end{tabular}}
\newcommand{\eaa}{\end{eqnarray*}}

\def\inbar{\,\vrule height1.5ex width.4pt depth0pt}
\def\IC{\relax\hbox{$\inbar\kern-.3em{\rm C}$}}
\def\IZ{\relax{\hbox{\cmss Z\kern-.4em Z}}}
\def\IR{{\hbox{{\rm I}\kern-.2em\hbox{\rm R}}}}

\def\IP{{\hbox{{\rm I}\kern-.2em\hbox{\rm P}}}}
\def\II{\hbox{{1}\kern-.25em\hbox{l}}}

\begin{document}

\title{Scalar and vector self-energies of heavy baryons in nuclear medium }
\date{\today}
\author{K. Azizi$^1$, N. Er$^2$, H. Sundu$^3$  \\
\textit{$^1$ Department of Physics, Do\v{g}u\c{s} University, Ac{\i}badem-Kad{\i}k\"oy, 34722 Istanbul, Turkey}\\
\textit{$^2$ Department of Physics, Abant \.{I}zzet Baysal University,
G\"olk\"oy Kamp\"us\"u, 14980 Bolu, Turkey}\\
\textit{$^3$ Department of Physics, Kocaeli University,  41380 \.{I}zmit, Turkey}}

\begin{abstract}
The in-medium sum rules are employed to calculate the shifts in the mass and residue as well as the scalar and vector self-energies of 
the heavy $\Lambda_Q, \Sigma_Q$ and $\Xi_Q$ baryons, with Q being $b$ or $c$ quark. The maximum shift in mass due to nuclear matter belongs 
to the  $\Sigma_c$ baryon and it is found to be $\Delta m_{\Sigma_{c}}=-936 ~ MeV$. In the case of residue, it is obtained that the residue 
 of  $\Sigma_b$ baryon is maximally affected by the nuclear medium with the shift $\Delta \lambda_{\Sigma_b} = -0.014 ~ GeV^3 $.
 The scalar and vector self-energies are found to be $\Sigma^{S}_{\Lambda_b} = 653 ~ MeV$, $\Sigma^{S}_{\Sigma_b} = -614 ~ MeV $,
 $\Sigma^{S}_{\Xi_b} = -17 ~ MeV $, $\Sigma^{S}_{\Lambda_c} = 272  ~ MeV $, $\Sigma^{S}_{\Sigma_c} = -936 ~ MeV $, $\Sigma^{S}_{\Xi_c} = -5 ~ MeV $ 
and $\Sigma^{\nu}_{\Lambda_b} = 436 \pm 148  ~ MeV $,  $\Sigma^{\nu}_{\Sigma_b} = 382 \pm 129 ~MeV $, $\Sigma^{\nu}_{\Xi_b} =15 \pm 5 ~ MeV$,  $\Sigma^{\nu}_{\Lambda_c} = 151 \pm 45 ~ MeV $, 
  $\Sigma^{\nu}_{\Sigma_c} = 486 \pm 144 ~ MeV $ and  $\Sigma^{\nu}_{\Xi_c} = 1.391 \pm 0.529 ~ MeV $. 
\end{abstract}

\maketitle



%

\section{Introduction}
The investigation of the in-medium properties of hadrons constituents  one of the main directions of the research in high energy and particle physics both theoretically and experimentally. Such studies can help us not only better understand the structure of the hot and dense astrophysical objects like the neutron stars and analyze the results of the heavy ion collision experiments, but also get valuable knowledge on the perturbative and non-perturbative aspects of QCD and the nature of the quark-gluon plasma as the new phase of matter. From the experimental side, the PANDA and CBM collaborations at FAIR aim to study the in-medium properties of not only the standard but also  the non-conventional exotic states newly discovered by different collaborations \cite{Prencipe:2015cgg, Biswas:2015paa, fair, panda, Friman, Lutz:2009ff, Giacosa:2015nwa}. 

From the theoretical side, it can be found many studies devoted to the in-medium properties of the light and heavy mesons as well as the light baryons (see for instance ~\cite{Drukarev:1988kd, Hatsuda:1990zj, Adami:1991js, Jin:1993fr, Jin:1994bh, Suzuki:2015est, Buchheim:2014uda, Buchheim:2015yyc, Azizi:2014yea, Azizi:2015ica, Hayashigaki:2000es, Hilger:2010zf, Wang:2011mj, Azizi:2014bba}).  In the heavy baryon sector, however, there are few works dedicated to the investigation of the spectroscopic parameters of the heavy baryons in nuclear medium \cite{Wang:2011hta, Wang:2011yj, Wang:2012xk}. In these studies, the authors use the Ioffe current to study some properties of heavy and doubly heavy baryons. 

In the present study, we use the interpolating currents with an arbitrary mixing parameter to investigate the shifts in the mass and pole residue of the heavy  $\Lambda_Q, \Sigma_Q$ and $\Xi_Q$ baryons due 
to the nuclear medium. We employ the in-medium QCD sum rule approach to calculate also the scalar and vector self-energies of those baryons by fixing the  mixing parameter entered  the interpolating currents according to the standard prescriptions. We shall note that we do not analyze the  heavy $\Omega_Q$ baryon here since the nuclear matter considered in the present 
work does not affect the parameters of the  $\Omega_Q$ baryon considerably as it does not contain the $u$ or $d$ quark. One may investigate the shifts on the parameters of this baryon in hyperonic and 
strange  matters or nuclear medium with strange component.  

This work is organized as follows. In section II, we obtain the in-medium QCD sum rules for mass and residue as well as the  scalar and vector self-energies of the heavy $\Lambda_Q, \Sigma_Q$ and $\Xi_Q$ baryons in nuclear matter. In section III, the numerical analyses of sum rules are performed and the results are compared with those of the vacuum and other predictions obtained via Ioffe currents in the literature. Last section  contains our concluding remarks. Some lengthy expressions obtained during calculations are moved to  the Appendix. 

\section{Mass, residue and self energies of heavy baryons in nuclear matter}
In order to calculate the mass, residue as well as the scalar and vector self-energies of the heavy baryons using the QCD sum rule method in nuclear medium, the first step is to construct an in-medium two point correlation function: 
\begin{eqnarray}\label{corre1}
\Pi = i \int d^4 x e^{i p \cdot x}  \langle  \psi_0| T [J_{B_{Q}} (x) \bar{J}_{B_{Q}}(0)] | \psi_0\rangle
\end{eqnarray}
where $p$ is the four momentum of the heavy baryon, $ | \psi_0\rangle$ is the nuclear matter ground state and  $J_{B_{Q}}$ is the interpolating current of the heavy
$B_Q = \Lambda_Q, \Sigma_Q, \Xi_Q$  baryons. These interpolating currents are given as
\begin{eqnarray} \label{current}
J_{\Sigma_Q}&=& -\frac{1}{\sqrt{2}} \epsilon^{abc} \Big\{\Big( q_{1}^{aT}CQ^b\Big)\gamma_5 q_2^c +\beta\Big(q_{1}^{aT}C\gamma_5 Q^b\Big) q_2^c \nonumber \\
&& -\Big[\Big(Q^{aT}Cq_{2}^{b}\Big) \gamma_5 q_1^c +\beta\Big(Q^{aT}C\gamma_5 q_{2}^{b}\Big) q_1^c \Big] \Big\} , \nonumber \\
J_{\Lambda_Q, \Xi_Q}&=&\frac{1}{\sqrt{6}} \epsilon^{abc} \Big\{2\Big( q_{1}^{aT}Cq_{2}^{b}\Big)\gamma_5 Q^c+2\beta\Big( q_{1}^{aT}C\gamma_5q_{2}^{b}\Big)Q^c \nonumber \\
&&+\Big( q_{1}^{aT}CQ^b\Big)\gamma_5 q_{2}^{c}+\beta\Big( q_{1}^{aT}C\gamma_5Q^b\Big) q_{2}^{c} \nonumber \\
&&+\Big( Q^{aT}Cq_{2}^{b}\Big)\gamma_5 q_{1}^{c} + \beta\Big( Q^{aT}C\gamma_5q_{2}^{b}\Big) q_{1}^{c} \Big\}
\end{eqnarray}
where $a, b, c$ are color indices and $C$ is the charge conjugation operator. The parameter $\beta$ is an arbitrary auxiliary parameter with $\beta=-1$ being corresponding to the Ioffe current, which  is obtained after a  Fierz transformation \cite{Thomas:2007gx} (for more details on the interpolating fields  see  \cite{Thomas:2007gx,Drukarev:2013kga,Leinweber:1994nm,Stein:1994zk}).
The quark fields $q_1$ and $q_2$ for different members are given in table I.
\begin{table}[h]

\renewcommand{\arraystretch}{1.3}
\addtolength{\arraycolsep}{-0.5pt}
\small
$$
\begin{array}{|l|c|c|c|}
\hline \hline 
&\Lambda_{b(c)}^{0(+)} &\Sigma_{b(c)}^{0(+)} &\Xi_{b(c)}^{0(+)}     \\  \hline
q_1 &u &u &u   \\
q_2 &d &d &s   \\
\hline \hline
\end{array}
$$
\caption{The quark flavors $q_1$ and $q_2$ for the baryons under consideration.}
\renewcommand{\arraystretch}{1}
\addtolength{\arraycolsep}{-1.0pt}
\end{table}

In accordance with the standard philosophy of  the method used, we shall calculate the aforesaid correlation function in hadronic and operator product expansion (OPE) sides.
 Matching the coefficients of different structures from these two sides through a dispersion relation leads to the sum rules for the different observables.
 In the hadronic side a complete set of heavy $B_Q$  baryon with the same quantum numbers as the interpolating currents is inserted into 
 the correlation function. After performing integral over four-x we get
\begin{eqnarray}\label{corre}
\Pi^{Had}&=&-\frac{\langle\psi_0|J_{B_Q}(0)|B_{Q}(p^*,s)\rangle\langle B_{Q}(p^*,s)|\bar{J}_{B_{Q}}(0)|\psi_0\rangle}{p^{*2}-m_{B_{Q}}^{*2}}  \nonumber \\
&&+ ...,
\end{eqnarray}
where $|B_{Q}(p^*,s)\rangle$ is  the heavy baryon state with spin $s$ and in-medium momentum $p^*$ and  $...$ represents the contributions of the higher states and the continuum. In the above equation, $m_{Q}^{*}$ is the modified mass of the heavy baryon in medium. The matrix elements seen in the above equation can be parametrized as
\begin{eqnarray}\label{intcur}
\langle\psi_0|J_{B_{Q}}(0)|B_Q(p^*,s)\rangle&=&\lambda_{B_{Q}}^{*}u_{B_Q}(p^*,s) \nonumber \\
\langle B_Q(p^*,s)|\bar{J}_{B_{Q}}(0)|\psi_0\rangle&=&\bar{\lambda}_{B_{Q}}^{*} \bar{u}_{B_Q}(p^*,s)
\end{eqnarray}
where $\lambda_{B_{Q}}^{*}$ is the modified residue or the coupling strengths of the heavy baryon to nuclear medium and  $u_{B_Q}(p^*,s)$ is the in-medium Dirac spinor. Using  Eq. (\ref{intcur}) in Eq. (\ref{corre}) 
and summing over the spins of the heavy baryon, the hadronic side of the in-medium correlation function can be written as
\begin{eqnarray}\label{corre2}
\Pi^{Had}&=&-\frac{\lambda_{B_{Q}}^{*2}(\!\not\!{p^*}+m_{B_Q}^{*})}{p^{*2}-m_{B_Q}^{*2}} + ...  = -\frac{\lambda_{B_{Q}}^{*2}}{\!\not\!p^{*}-m_{B_Q}^{*}} + ... \nonumber \\
&=&-\frac{\lambda_{B_{Q}}^{*2}}{(p^{\mu}_{B_Q}-\Sigma_{\nu B_Q}^{\mu})\gamma_\mu-(m_{B_Q}+\Sigma^{S}_{B_Q})}  + ...,
\end{eqnarray}
where $\Sigma^{\mu}_{\nu B_{Q}}$ and $\Sigma^{S}_{B_{Q}}$ are the vector and  scalar self-energies of the heavy baryon in nuclear matter, respectively \cite{Cohen:1991js}. The vector self energy  can be written as 
\begin{equation}\label{sigma1}
\Sigma_{\nu B_{Q}}^{\mu}=\Sigma^{\nu}_{B_{Q}} u^\mu+\Sigma'_{\nu B_{Q}}p_{B_{Q}}^\mu,
\end{equation}
where $ u^\mu$ is the four velocity of the nuclear medium and we ignore $\Sigma'_{\nu B_{Q}}$ because of its small value \cite{Cohen:1994wm} . The four-velocity of the nuclear matter creates extra structures to the correlation function compared to  the vacuum QCD sum rules. The calculations are performed in the
rest frame of the nuclear medium,  i.e. $u^\mu=(1,0)$. After the substitution of Eq. (\ref{sigma1}) into Eq. (\ref{corre2}), 
we get the hadronic side of the correlation function  in terms of three different structures  as  
\begin{eqnarray}
\Pi^{Had}&=&\Pi^{Had}_{p}(p^2,p_0)\!\not\!{p}+\Pi^{Had}_{u}(p^2,p_0)\!\not\!{u} \nonumber \\
&+&\Pi^{Had}_{S}(p^2,p_0)I+ ...,
\end{eqnarray}
here $p_0$ is the energy of the quasi-particle, $I$ is the unit matrix and
\begin{eqnarray}
\Pi^{Had}_{p}(p^2,p_0)&=&-\lambda_{B_{Q}}^{*2}\frac{1}{p^2-\mu_{B_Q}^2},\nonumber \\
\Pi^{Had}_{u}(p^2,p_0)&=&+\lambda_{B_{Q}}^{*2}\frac{\Sigma^{\nu }_{B_{Q}}}{p^2-\mu_{B_Q}^2},
\nonumber \\
\Pi^{Had}_{S}(p^2,p_0)&=&-\lambda_{B_{Q}}^{*2}\frac{m_{B_Q}^*}{p^2-\mu_{B_Q}^2} ,
\end{eqnarray}
where $m_{B_{Q}}^*=m_{B_{Q}}+\Sigma^{S}_{B_{Q}}$ and
$\mu_{B_{Q}}^2=m_{B_{Q}}^{*2}-\Sigma^{\nu 2}_{B_{Q}}+2p_{0}\Sigma^{\nu}_{B_{Q}}$. After  applying the Borel transformation with respect to $p^2$, we obtain
\begin{eqnarray}\label{3eqn6unknown}
\hat{B}\Pi^{Had}_{p}(p^2,p_0)&=&\lambda_{B_{Q}}^{*2}e^{-\mu_{B_Q}^2/M^2}, \nonumber \\
\hat{B}\Pi^{Had}_{u}(p^2,p_0)&=&-\lambda_{B_{Q}}^{*2}\Sigma_{\nu B_Q} e^{-\mu_{B_Q}^2/M^2},
\nonumber \\
\hat{B}\Pi^{Had}_{S}(p^2,p_0)&=&\lambda_{B_Q}^{*2} m_{B_Q}^* e^{-\mu_{B_Q}^2/M^2},
\end{eqnarray}
where $M^2$ is the Borel mass parameter to be fixed in  next section. 

The QCD representation of the correlation function is derived in the deep Euclidean region. The correlator in this side is also decomposed in terms of the selected structures as 
\begin{eqnarray}
\Pi^{QCD}&=&\Pi_{p}^{QCD} (p^2, p_0)\!\not\!{p}+\Pi_{u}^{QCD} (p^2, p_0)\!\not\!{u} \nonumber \\
&+&\Pi_{S}^{QCD} (p^2, p_0) I.
\end{eqnarray}
To calculate the correlation function in QCD side, we insert the interpolating currents (\ref{current}) into the correlation function (\ref{corre1}) and perform the contractions of the quark pair using the Wick's theorem. As a result, we get expressions presented in the Appendix for different particles in terms of the heavy and light quark propagators. 

To proceed, we need to know the expressions for the quark propagators in the coordinate space.
In  the fixed point gauge, the light and heavy quark propagators are chosen as  
\begin{eqnarray}
S_q^{ij}(x)&=&
\frac{i}{2\pi^2}\delta^{ij}\frac{1}{(x^2)^2}\not\!x
-\frac{m_q }{ 4\pi^2} \delta^ { ij } \frac { 1}{x^2} + \chi^i_q(x)\bar{\chi}^j_q(0) \nonumber \\
&-&\frac{ig_s}{32\pi^2}F_{\mu\nu}^A(0)t^{ij,A}\frac{1}{x^2}[\not\!x\sigma^{\mu\nu}+\sigma^{\mu\nu}\not\!x] +\cdots \, ,\nonumber\\
S_Q^{ij}(x)&=&\frac{i}{(2\pi)^4}\int d^4k e^{-ik \cdot x} \left\{\frac{\delta_{ij}}{\!\not\!{k}-m_Q}\right.\nonumber\\
&&\left.-\frac{g_sF_{\mu\nu}^A(0)t^{ij,A}}{4}\frac{\sigma_{\mu\nu}(\!\not\!{k}+m_Q)+(\!\not\!{k}+m_Q)
\sigma_{\mu\nu}}{(k^2-m_Q^2)^2}\right.\nonumber\\
&&\left.+\frac{\pi^2}{3} \langle \frac{\alpha_sGG}{\pi}\rangle\delta_{ij}m_Q \frac{k^2+m_Q\!\not\!{k}}{(k^2-m_Q^2)^4}+\cdots\right\} \, ,
\end{eqnarray}
where $\chi^i_q$ and $\bar{\chi}^j_q$ are the Grassmann background quark fields,
$F_{\mu\nu}^A$ are classical background gluon fields, and $t^{ij,A}=\frac{\lambda ^{ij,A}}{2}$ with $
\lambda ^{ij, A}$ being  the standard Gell-Mann matrices. The quark, gluon and mixed condensates are defined in terms of different operators in nuclear matter. To avoid from redundancy, we do not present their explicit expressions here, but refer the readers to Refs.  \cite{Cohen:1994wm, Azizi:2014yea} for more details. 

The QCD side of the correlation function in the Borel scheme can be written  in terms of the  perturbative and non-perturbative parts as
\begin{eqnarray} \label{OPE}
\widehat{\textbf{B}}\Pi^{QCD}_{p, u, S}(p^2, p_0)&=&\widehat{\textbf{B}}\Pi_{p, u, S}^{Pert}(p^2, p_0)+\widehat{\textbf{B}}\Pi_{p, u, S}^{qq}(p^2, p_0)\nonumber \\
&&+\widehat{\textbf{B}}\Pi_{p, u, S}^{GG}(p^2, p_0)+\widehat{\textbf{B}}\Pi_{p, u, S}^{qGq}(p^2, p_0) \nonumber \\&&+\widehat{\textbf{B}}\Pi_{p, u, S}^{qqqq}(p^2, p_0),
\end{eqnarray}
where $ Pert $ represents  the perturbative part and  the upper indices  $ qq $,  $ GG $, $ qGq $ and $ qqqq $ denote the contributions of two-quark, two-gluon, mixed and four-quark condensates, respectively.  Writing each function in terms of the even and odd parts as $ \widehat{\textbf{B}}\Pi^{i}_{j}(p^2, p_0)=\widehat{\textbf{B}}\Pi^{i}_{j,E}(p^2, p_0)+p_0 \widehat{\textbf{B}}\Pi^{i}_{j, O}(p^2, p_0) $, for the structure $\!\not\!{p}$ and $\Lambda_Q$ baryon we obtain
\begin{widetext}
\begin{eqnarray}
\widehat{\textbf{B}}\Pi_{p, E}^{Pert}(p^2, p_0) &=&   \frac{1}{2048 \pi^4} (5+2\beta + 5 \beta^2)  \int^{s_0}_{m_{Q}^2} \Big \{  m^{8}_{Q}-8m^{6}_{Q} s + 8 m^{2}_{Q} s^3 -s^4 +12 m^{4}_{Q} s^2 Log[m^{2}_{Q}/s]  \Big \} \frac{e^{-\frac{s}{M^2}}}{ s^2}ds, 
\nonumber \\
\widehat{\textbf{B}}\Pi_{p, O}^{Pert}(p^2, p_0) &=&  0,
\nonumber \\
\widehat{\textbf{B}}\Pi_{p, E}^{qq}(p^2, p_0)&=&  \frac{1}{1728 \pi^2} \int^{s_0}_{m_{Q}^2}  \Big\{-(m^{2}_{Q}-s)^2 \Big [\Big(9(1+4\beta-5\beta^{2})m_Q s+4(19+7\beta+19\beta^2)m_q(2m_Q^{2}+s)\Big ) \langle \bar{u}u \rangle_{\rho_N} \nonumber \\
&& - \Big(2i(17+2\beta+17\beta^2)(2 m^{2}_{Q}+s)\Big) \langle u^{\dag} iD_0 u\rangle_{\rho_N}   \nonumber \\
&& +\Big(9(1+4\beta-5\beta^2)m_Q s+4(19+7\beta+19\beta^2)m_q(2m_Q^{2}+s)\Big ) \langle \bar{d}d \rangle_{\rho_N} \nonumber \\
&& - \Big(2i(17+2\beta+17\beta^2)(2 m^{2}_{Q}+s)\Big) \langle d^{\dag} iD_0 d\rangle_{\rho_N}  \Big] \Big\}    \frac{e^{-\frac{s}{M^2}}}{ s^3}ds,
\nonumber \\
\widehat{\textbf{B}}\Pi_{p, O}^{qq}(p^2, p_0)&=&  \frac{1}{288 \pi^2} \int^{s_0}_{m_{Q}^2}  \Big\{\Big( (7+4\beta+7\beta^2)(m^{2}_{Q}-s)^2(2 m^{2}_{Q}+s)\Big)\langle u^{\dag} u\rangle_{\rho_N} \nonumber \\
&&+\Big( (7+4\beta+7\beta^2)(m^{2}_{Q}-s)^2(2 m^{2}_{Q}+s)\Big)\langle d^{\dag} d\rangle_{\rho_N}  \Big\}    \frac{e^{-\frac{s}{M^2}}}{ s^3}ds,
 \nonumber \\
\widehat{\textbf{B}}\Pi_{p, E}^{GG}(p^2, p_0)&=& \frac{1}{3072 \pi^2 } (1+\beta)^2 \int^{s_0}_{m_{Q}^2} \Big \{  (m^{2}_{Q} -s) \Big [2 m^{4}_{Q} +11 m^{2}_{Q}s +3 s^2 \Big ] \langle \frac{\alpha_s}{\pi} G^{2}\rangle_{\rho_N} \Big \}\frac{e^{-\frac{s}{M^2}}}{ s^3}ds,
\nonumber \\
\widehat{\textbf{B}}\Pi_{p, O}^{GG}(p^2, p_0)&=&0,
 \nonumber \\
\widehat{\textbf{B}}\Pi_{p,E(O)}^{qGq}(p^2, p_0)&=& 0,
\end{eqnarray}
\end{widetext}
where $ s_0 $ is the continuum threshold and we do not show the even and odd parts of $\widehat{\textbf{B}}\Pi_{p, u, S}^{qqqq}(p^2, p_0)$ here because of their very lengthy expressions.  The four-quark condensate is poorly known in the nuclear medium.  Hence, by the help of a density dependent factor, it is factorized into multiplication of two two-quark condensates. The value of this factor is determined by examination of its  effect  on predictions from QCD sum rules (for details see \cite{Thomas:2007gx}) . As the impact of this factor is very weak in the case of heavy baryons we use   the naive factorizations \cite{Cohen:1994wm}
\begin{eqnarray}
\langle q_{a\alpha}\bar{q}_{b\beta}q_{c\gamma}\bar{q}_{d\delta}\rangle_{\rho_N}&\simeq &
\langle q_{a\alpha}\bar{q}_{b\beta}\rangle_{\rho_N}\langle q_{c\gamma}\bar{q}_{d\delta}\rangle_{\rho_N}
\nonumber \\
&-&\langle q_{a\alpha}\bar{q}_{d\delta}\rangle_{\rho_N}\langle q_{c\gamma}\bar{q}_{b\beta}\rangle_{\rho_N},
\end{eqnarray}
in the case of same quarks and 
\begin{eqnarray}
\langle q_{a\alpha}\bar{q}_{b\beta}q'_{c\gamma}\bar{q'}_{d\delta}\rangle_{\rho_N}&\simeq &
\langle q_{a\alpha}\bar{q}_{b\beta}\rangle_{\rho_N}\langle q'_{c\gamma}\bar{q'}_{d\delta}\rangle_{\rho_N}
\end{eqnarray}
for different quarks
as well as  the expansion \cite{Cohen:1994wm},
\begin{eqnarray} \label{ }
\langle
q_{a\alpha}(x)\bar{q}_{b\beta}(0)\rangle_{\rho_N}&=&-\frac{\delta_{ab}}{12}
\Bigg
[\Bigg(\langle\bar{q}q\rangle_{\rho_N}+x^{\mu}\langle\bar{q}D_{\mu}q\rangle_{
\rho_N}
 \nonumber \\
&+&\frac{1}{2}x^{\mu}x^{\nu}\langle\bar{q}D_{\mu}D_{\nu}q\rangle_{\rho_N}
+...\Bigg)\delta_{\alpha\beta}\nonumber \\
&&+\Bigg(\langle\bar{q}\gamma_{\lambda}q\rangle_{\rho_N}+x^{\mu}\langle\bar{q}
\gamma_{\lambda}D_{\mu} q\rangle_{\rho_N}
 \nonumber \\
&+&\frac{1}{2}x^{\mu}x^{\nu}\langle\bar{q}\gamma_{\lambda}D_{\mu}D_{\nu}
q\rangle_{\rho_N}
+...\Bigg)\gamma^{\lambda}_{\alpha\beta} \Bigg],\nonumber \\
\end{eqnarray}
in terms of the operators having $ 3,4 $ and $ 5 $ mass dimensions in nuclear matter. We ignore
from the contributions of the more than four-quark and two-gluon operators as their values are unknown at finite density (for more details see for instance \cite{Cohen:1994wm}).

As a final goal, the coefficients of three structures from both hadronic and QCD sides are matched. This leads to the following sum rules: 
\begin{eqnarray}\label{3eqn8unknownOPE}
\lambda_{B_{Q}}^{*2}e^{-\mu_{B_Q}^2/M^2} &=& \widehat{\textbf{B}}\Pi^{QCD}_{p}, \nonumber \\
-\lambda_{B_{Q}}^{*2}\Sigma^{\nu }_{B_{Q}} e^{-\mu_{B_Q}^2/M^2} &=&\widehat{\textbf{B}} \Pi^{QCD}_{u},
\nonumber \\
\lambda_{B_{Q}}^{*2} m_{B_Q}^* e^{-\mu_{B_Q}^2/M^2} &=& \widehat{\textbf{B}}\Pi^{QCD}_{S}.
\end{eqnarray}
By solving these equations and an extra equation, that is obtained by applying a derivative to both sides of the first equation with respect to $ -1/M^2 $, simultaneously we find the parameters   $\lambda_{B_{Q}}^{*}, m_{B_Q}^{*}, \Sigma^{\nu}_{B_{Q}}$ and $\mu_{B_Q}$ as 
 
\begin{eqnarray}
\mu_{B_Q}^2 (s_0, M^2, \beta) &=& \frac{d \widehat{\textbf{B}}\Pi^{QCD}_{p}/d(-1/M^2)}{\widehat{\textbf{B}}\Pi^{QCD}_{p}}, \nonumber \\
\lambda_{B_{Q}}^{*2} (s_0, M^2, \beta)&=& e^{\mu_{B_Q}^2/M^2} \widehat{\textbf{B}}\Pi^{QCD}_{p}, \nonumber \\
 \Sigma^{\nu}_{ B_Q} (s_0, M^2, \beta) &=& -\frac{\widehat{\textbf{B}}\Pi^{QCD}_{u}}{\widehat{\textbf{B}}\Pi^{QCD}_{p}}, \nonumber \\
 m_{B_q}^{*} (s_0, M^2, \beta) &=& \frac{\widehat{\textbf{B}}\Pi^{QCD}_{S}}{\widehat{\textbf{B}}\Pi^{QCD}_{p}}.
 \end{eqnarray}
 Alternatively, we can use the previously introduced relation,
 \begin{eqnarray}
 m_{B_q}^{*2} (s_0, M^2, \beta) &=& \mu^{2}_{B_Q}+ \Sigma^{\nu 2}_{B_{Q}} - 2 p_0  \Sigma^{\nu }_{B_{Q}},
\end{eqnarray}
to find the mass of the baryons under consideration at finite density. 

Here we should remark that by the method used for the in-medium heavy baryons in the present study including the Borel transformation with respect to $ p^2 $ and the chosen continuum threshold we automatically extracted the contributions of the particles with positive energy excitations.  This can be considered  as an alternative to the canonical approach to in-medium QCD sum rules for nucleons  \cite{Cohen:1994wm,Furnstahl:1995nd}, which uses Cauchy's theorem in the $ p_0 $ plane with a cut along the full $  Im [p_0] = 0$ axis thus including necessarily positive and negative frequencies at fixed $ \vec p  $. For the light baryons especially the nucleons which contain exactly the same quark ingredients  with the medium one shall consider both the particle and antiparticle contributions at fixed three-momentum of the particles and separate the contributions of the positive frequency particles by the help of an appropriate weight function (for details see also \cite{Thomas:2007gx}).
\section{Numerical Results and Discussion}
The numerical analyses of the QCD sum rules for the mass, pole residue, scalar and vector self-energies of the heavy $\Lambda_Q, \Sigma_Q$ and $\Xi_Q$ in nuclear medium, require the values of quark and baryon masses, nuclear matter density, different in-medium quark, gluon and mixed condensates, etc. Their numerical values are collected in table II.
\begin{widetext}

\begin{table}[ht!]
\centering
\begin{tabular}{ll}
\hline \hline
   Input parameters  &  Values    
           \\
\hline \hline
$p_0   $          &  $m_{B_Q} $      \\
$ m_{u}   $ ; $ m_{d}   $ ;  $ m_{s}  $  &  $2.3_{-0.5}^{0.7}  $ $MeV$     ; $4.83_{-0.3}^{0.5}  $ $MeV$ ;   $95 \pm 5  $ $MeV$   \cite{PDG}      \\
$ m_{b}   $ ; $ m_{c}   $     &  $4.78 \pm  $ $GeV$     ; $1.275 \pm 0.025  $ $GeV$    \cite{PDG}      \\
$ m_{\Lambda_{b}} $; $ m_{\Lambda_{c}}  $  &  $ 5619.51 \pm 0.23 $ $MeV$ ;  $ 2286.46 \pm 0.14 $ $MeV$  \cite{PDG}  \\
$ m_{\Sigma_{b}} $; $ m_{\Sigma_{c}}  $  &  $ 5811.3 \pm 1.9$ $MeV$  ;  $ 2452.9 \pm 0.4 $ $MeV$  \cite{PDG}  \\
$ m_{\Xi_b} $; $ m_{\Xi_c}  $  &  $  5791.8 \pm 0.5$ $MeV$ ;  $ 2467.93^{+0.28}_{-0.40}$ $MeV$  \cite{PDG}  \\
$ \rho_{N}     $          &  $(0.11)^3  $ $GeV^3$   \cite{Cohen:1991nk}     \\
$ \langle q^{\dag}q\rangle_{\rho_N}    $ ; $ \langle s^{\dag}s\rangle_{\rho_N}    $         &  $\frac{3}{2}\rho_{N}$   ; 0     \cite{Cohen:1991nk}    \\
$ \langle\bar{q}q\rangle_{0} $  ; $ \langle\bar{s}s\rangle_{0} $  &  $ (-0.241)^3    $ $GeV^3$ ; 0.8 $ \langle\bar{q}q\rangle_{0} $    \cite{Belyaev:1982cd, Ioffe:2005ym}        \\
$ m_{q}      $          &  $0.5(m_{u}+m_{d})$        \cite{Cohen:1991nk}      \\
$ \sigma_{N} $  ;   $ \sigma_{N_0} $      &  $0.045 ~  $GeV$ $ ; $0.035 ~  $GeV$ $  \cite{Cohen:1991nk}   \\
$y$ & $0.04\pm0.02$ \cite{Thomas1}; $0.066\pm0.011\pm0.002$ \cite{Dinter:2011za} \\
$  \langle\bar{q}q\rangle_{\rho_N}  $ ;  $  \langle\bar{s}s\rangle_{\rho_N}  $  &  $ \langle\bar{q}q\rangle_{0}+\frac{\sigma_{N}}{2m_{q}}\rho_{N}$ 
; $ \langle\bar{s}s\rangle_{0}+y\frac{\sigma_{N}}{2m_{q}}\rho_{N}$   \cite{Cohen:1991nk}             \\
$  \langle q^{\dag}g_{s}\sigma Gq\rangle_{\rho_N}  $ ;  $  \langle s^{\dag}g_{s}\sigma Gs\rangle_{\rho_N} $ &  $ -0.33~GeV^2 \rho_{N}$ ; $ -y 0.33~GeV^2 \rho_{N}$ \cite{Cohen:1991nk} \\
$  \langle q^{\dag}iD_{0}q\rangle_{\rho_N}  $    ; $  \langle s^{\dag}iD_{0}s\rangle_{\rho_N}  $      &  $0.18 ~GeV \rho_{N}$  ;$ \frac{m_s \langle\bar{s}s\rangle_{\rho_N}}{4}
+0.02 ~GeV \rho_N$      \cite{Cohen:1991nk}    \\
$  \langle\bar{q}iD_{0}q\rangle_{\rho_N}  $    ; $  \langle\bar{s}iD_{0}s\rangle_{\rho_N}  $      &  $\frac{3}{2} m_q \rho_{N}\simeq0 $ ; 0 \cite{Cohen:1991nk} \\
$  m_{0}^{2}  $          &  $ 0.8~GeV^2$   \cite{Belyaev:1982cd, Ioffe:2005ym}          \\
$   \langle\bar{q}g_{s}\sigma Gq\rangle_{0} $ ; $   \langle\bar{s}g_{s}\sigma Gs\rangle_{0} $  &  $m_{0}^{2}\langle\bar{q}q\rangle_{0} $ ; $m_{0}^{2}\langle\bar{s}s\rangle_{0} $ \\
$  \langle\bar{q}g_{s}\sigma Gq\rangle_{\rho_N}  $  ;$  \langle\bar{s}g_{s}\sigma Gs\rangle_{\rho_N}  $ &  $\langle\bar{q}g_{s}\sigma Gq\rangle_{0}+3~GeV^2\rho_{N} $ ;
$\langle\bar{s}g_{s}\sigma Gs\rangle_{0}+3 y ~GeV^2\rho_{N} $ \cite{Cohen:1991nk} \\
$ \langle  \bar{q}iD_{0}iD_{0}q\rangle_{\rho_N} $  ;$\langle \bar{s}iD_{0}iD_{0}s\rangle_{\rho_N} $  &  $ 0.3~GeV^2\rho_{N}-\frac{1}{8}\langle\bar{q}g_{s}\sigma Gq\rangle_{\rho_N}$ ; $ 0.3 y ~GeV^2\rho_{N}-\frac{1}{8}\langle\bar{s}g_{s}\sigma Gs\rangle_{\rho_N}$ \cite{Cohen:1991nk} \\
$  \langle q^{\dag}iD_{0}iD_{0}q\rangle_{\rho_N}  $ ;$  \langle s^{\dag}iD_{0}iD_{0}s\rangle_{\rho_N}$ & $0.031~GeV^2\rho_{N}-\frac{1}{12}\langle q^{\dag}g_{s}\sigma Gq\rangle_{\rho_N} $; $0.031 y~GeV^2\rho_{N}-\frac{1}{12}\langle s^{\dag}g_{s}\sigma Gs\rangle_{\rho_N} $\cite{Cohen:1991nk} \\
$\langle \frac{\alpha_s}{\pi} G^{2}\rangle_{0}$ & $(0.33\pm0.04)^4~GeV^4$  \cite{Belyaev:1982cd, Ioffe:2005ym} \\
$\langle \frac{\alpha_s}{\pi} G^{2}\rangle_{\rho_N}$ & $\langle \frac{\alpha_s}{\pi} G^{2}\rangle_{0}-0.65~GeV \rho_N$ \cite{Cohen:1991nk} \\
 \hline \hline
\end{tabular}
\caption{Numerical values for input parameters. Note that we use the average value $y=0.05$ to perform the numerical analyses.}
\end{table}

\end{widetext}Note that we use the average value $y=0.05$ to perform the numerical analyses.

Besides the above input parameters, the sum rules obtained for the mass, residue and self energies of heavy baryons contain three more auxiliary parameters that should be fixed: the Borel parameter $M^2$, the continuum threshold $s_0$ and the mixing parameter $\beta$. The standard way in determining the working regions of these parameters is that the physical observables demonstrate weak dependence on them. The continuum threshold is not totally arbitrary but it depends on the energy of the first excited state with the same quantum numbers. This parameter is related to the beginning of the continuum in the channels under consideration. If the ground state mass is given by $ m_{B_Q} $, $ \sqrt{s_0}- m_{B_Q}$ is the energy needed to excite the particle to its first excited state. According to the standard prescriptions and considering the masses of the first excited states in the channels under study, we choose this energy in the interval $ 0.1~GeV-0.4~GeV$, which leads to the following intervals for the b-baryons under consideration:
\begin{eqnarray}
32.7 ~ GeV^2  \leqslant s^{\Lambda_b}_0 \leqslant 36.2 ~ GeV^2, \nonumber \\
34.9 ~ GeV^2  \leqslant s^{\Sigma_b}_0 \leqslant 38.6 ~ GeV^2, \nonumber \\
34.7 ~ GeV^2  \leqslant s^{\Xi_b}_0 \leqslant 38.3 ~ GeV^2
\end{eqnarray}
for the $c$-baryons we get 
\begin{eqnarray}
5.7 ~ GeV^2  \leqslant s^{\Lambda_c}_0 \leqslant 7.2 ~ GeV^2 ,\nonumber \\
6.5 ~ GeV^2  \leqslant s^{\Sigma_c}_0 \leqslant 8.1 ~ GeV^2,\nonumber \\
6.6 ~ GeV^2  \leqslant s^{\Xi_c}_0 \leqslant 8.2 ~ GeV^2.
\end{eqnarray}
Our analyses show that in these intervals, the dependence of the physical parameters on the continuum threshold is weak. The standard criteria in calculating the Borel window are that not only the pole contribution exceeds the contributions of the higher states and continuum but also the series of
 OPE converge, i.e. the higher operators contribute less to the total integral compared to the lower operators and the perturbative contribution overcomes to the nonperturbative ones. These criteria lead to the following intervals: 
\begin{eqnarray}
5~ GeV^2 \leqslant ~ M^2 \leqslant 8 ~ GeV^2 ~~~~~ \textrm{for} ~~ \Lambda_b, \Sigma_b, \Xi_b,
\end{eqnarray}
and
\begin{eqnarray}
3~ GeV^2 \leqslant ~ M^2 \leqslant 6 ~ GeV^2 ~~~~~  \textrm{for} ~~ \Lambda_c, \Sigma_c, \Xi_c.
\end{eqnarray}
The pole dominance together with the aforesaid intervals for the Borel and threshold parameters lead to the following windows to the  parameter $\beta$ 
\begin{eqnarray}
-0.6 \leqslant x \leqslant -0.4 ~~~ \textrm{and} ~~~ 0.4 \leqslant x \leqslant 0.6.
\end{eqnarray}
where we used $x=\cos\theta$ with $\theta=\tan^{-1} \beta$ notation to explore the whole region from $- \infty$ to $+\infty$ by varying $x$ in the interval $[-1,1]$. Note that the Ioffe current with $x=-0.71$ remains out of the reliable regions in our calculations. Our numerical calculations show that in the above regions the results relatively show weak dependence on the $x$ or $\beta$. 

Now, we proceed to numerically analyze the sum rules obtained for the physical observables using these working windows and the values of other input parameters. 
To this end, first of all, in order to show how the OPE converges in our calculations we compare the variations of perturbative part, two-quark condensate, two-gluon condensate, mixed condensate and four-quark condensate with approximations in Eqs. (14) and (15) for instance in $ \Lambda_b $ channel and $\!\not\!{p}$ structure with respect to $ M^{2} $ at average values of other auxiliary parameters in figure 1.  
\begin{figure}[h!]
\label{fig1}
\centering
\begin{tabular}{ccc}
\epsfig{file=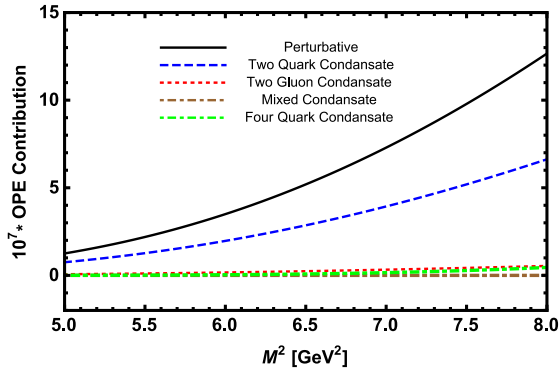,width=0.85\linewidth,clip=} 
\end{tabular}
\caption{Variations of perturbative part, two-quark condensate, two-gluon condensate, mixed condensate and four-quark condensate in $ \Lambda_b $ channel and $\!\not\!{p}$ structure with respect to $M^2$ at average values of $s_0$ and $x$. }
\end{figure}
From this figure, we see that the OPE nicely converges, i.e., the perturbative part exceeds the nonperturbative contributions and  the contributions  reduce with increasing the dimension. Note that as is also clear from Eq. (13) the contribution of mixed condensate is exactly zero since the terms containing the mixed condensate have no imaginary parts and do not contribute to the spectral density. We also see that the four-quark condensate has least contribution to the sum rules and the approximations (14) and (15) seem reasonable.  Similar results are obtained for other channels and structures.  We plot the quantities under consideration, i.e. masses, residues, the vector self-energies with respect to $M^2$ at average values of the threshold 
and mixing parameters in figures 2-10 for both the nuclear medium and vacuum. Only in the case of vector self energies we depict their variations with respect to the Borel parameter 
for different particles in medium since they do not exist in the vacuum. We shall also remark that we calculate the scalar self energies via the shifts in 
masses compared to their vacuum values, hence we do not plot their variations with respect to $M^2$, separately. First of all, we see that these figures 
depict considerable shifts on the values of observables due to nuclear matter when we compare them with their vacuum values. 
The next issue that should be emphasized is: the physical observables under consideration overall show weak dependence on the Borel parameter both in vacuum and nuclear medium in the selected windows. 
Extracted from the numerical calculations, we present the average numerical results for different quantities for both the nuclear matter and vacuum and also both the $b$ and $c$-baryons 
in tables III-V. The quoted errors in the values are due to the uncertainties coming from the calculations of the working regions of auxiliary parameters, errors of different input parameters as well as those related to different approximations used in the analyses.    We present the existing predictions for parameters in $\Lambda_Q$ and $\Sigma_Q$ channels obtained via Ioffe current \cite{Wang:2011hta, Wang:2011yj}  in tables III and IV as well. 

\begin{widetext}

\begin{figure}[h!]
\label{fig1}
\centering
\begin{tabular}{ccc}
\epsfig{file=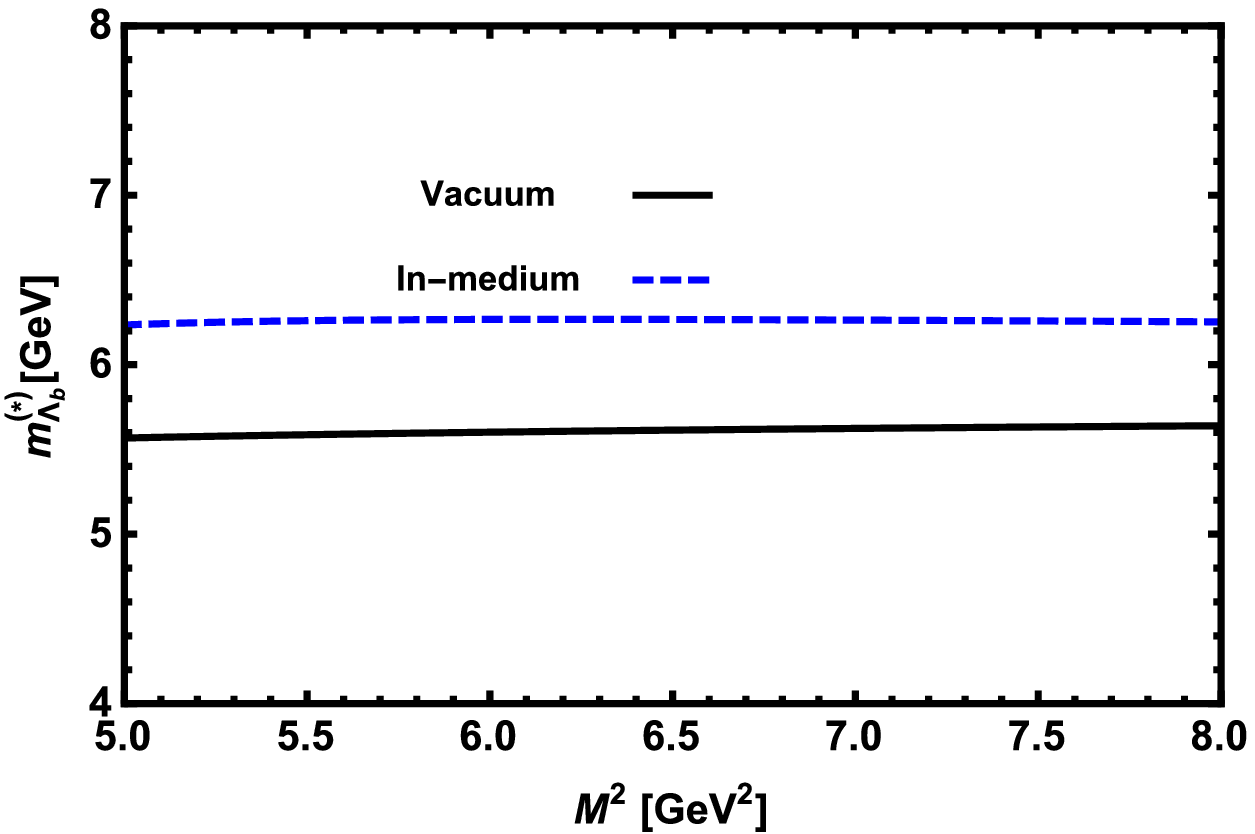,width=0.45\linewidth,clip=} &
\epsfig{file=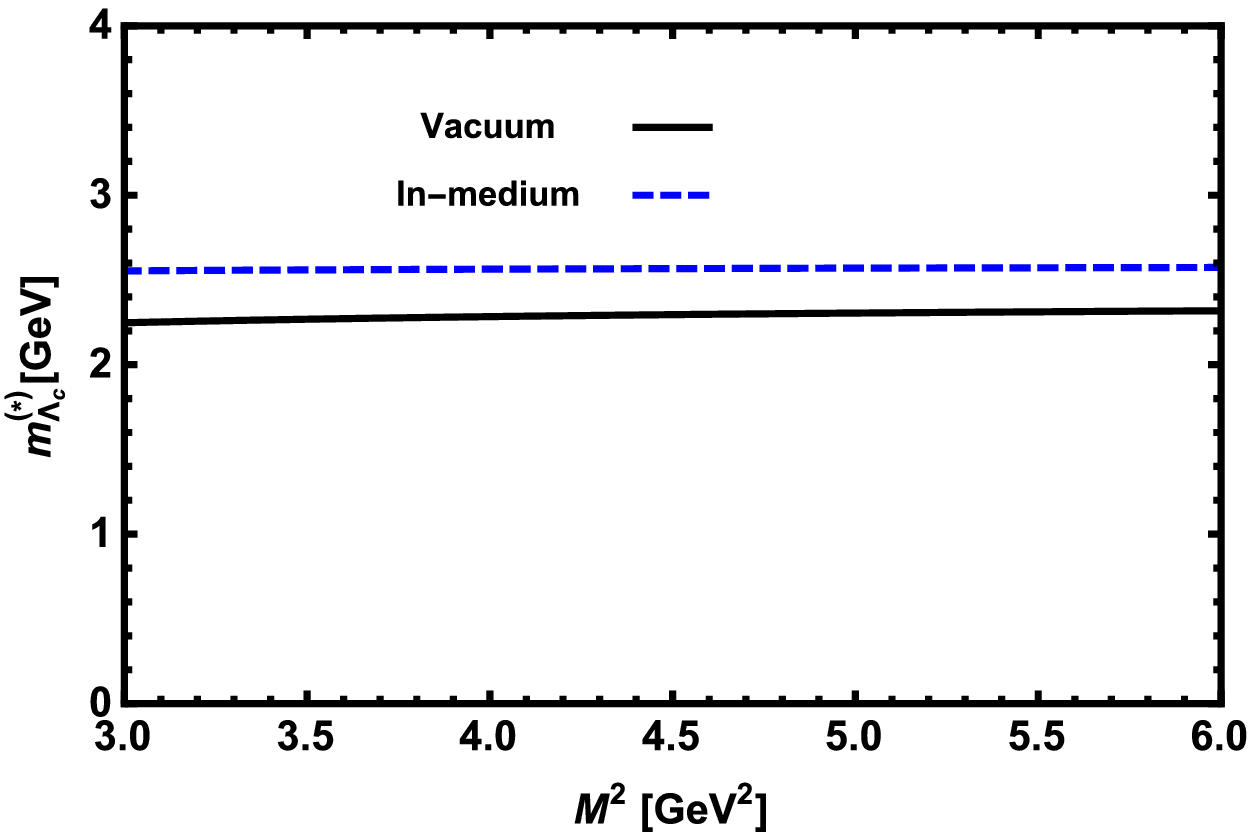,width=0.45\linewidth,clip=} 
\end{tabular}
\caption{The mass of the $\Lambda_b$  baryon (left panel) and  the $\Lambda_c$  baryon (right panel) versus  $M^2$  in vacuum and nuclear medium at average values of $s_0$ and $x$. }
\end{figure}

\begin{figure}[h!]
\label{fig1}
\centering
\begin{tabular}{ccc}
\epsfig{file=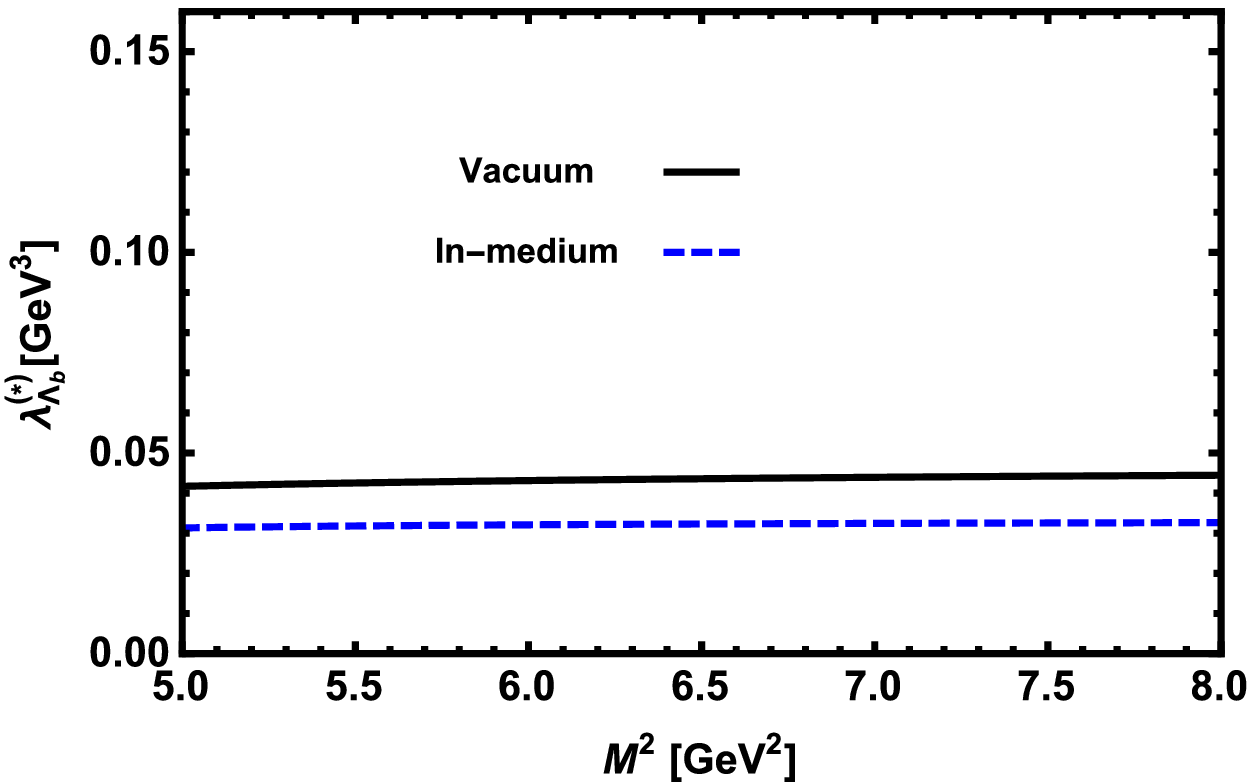,width=0.45\linewidth,clip=} &
\epsfig{file=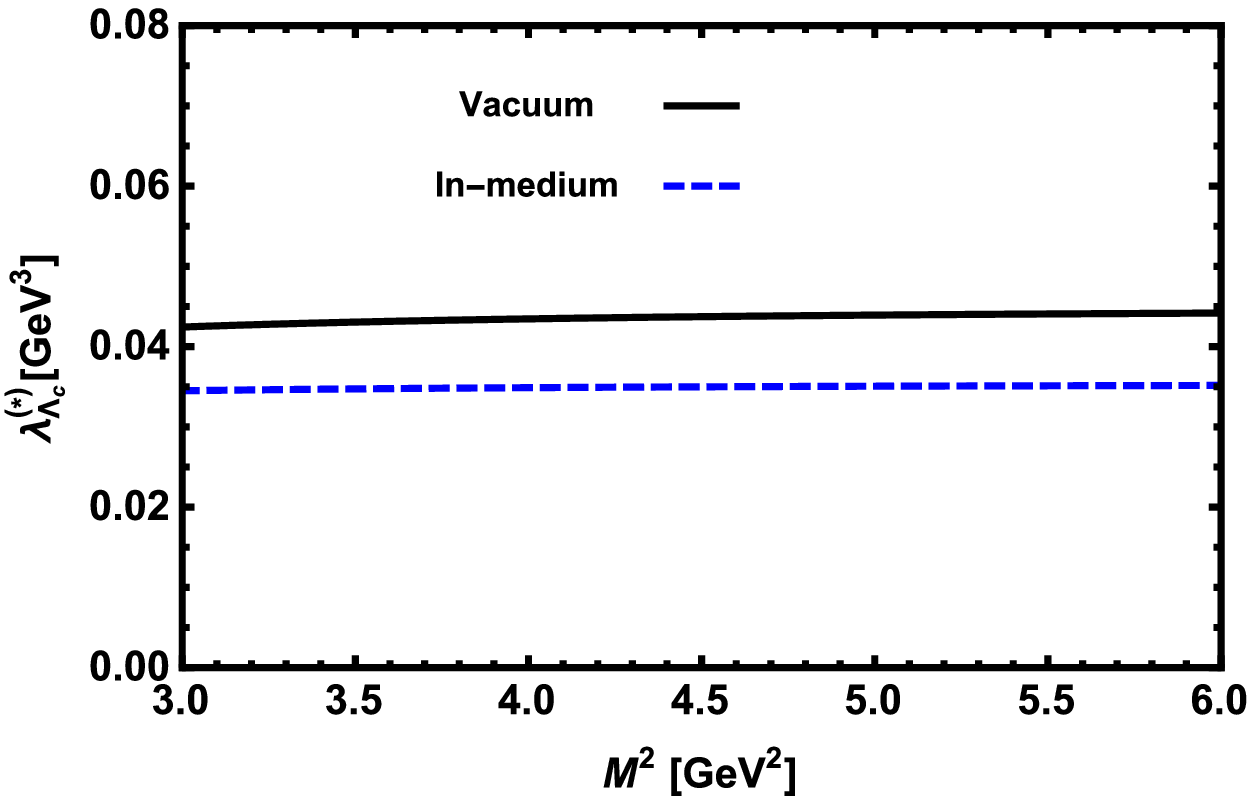,width=0.45\linewidth,clip=} 
\end{tabular}
\caption{The residue of the $\Lambda_b$  baryon (left panel) and  the $\Lambda_c$  baryon (right panel) versus  $M^2$  in vacuum and nuclear medium at average values of $s_0$ and $x$. }
\end{figure}

\begin{figure}[h!]
\label{fig1}
\centering
\begin{tabular}{ccc}
\epsfig{file=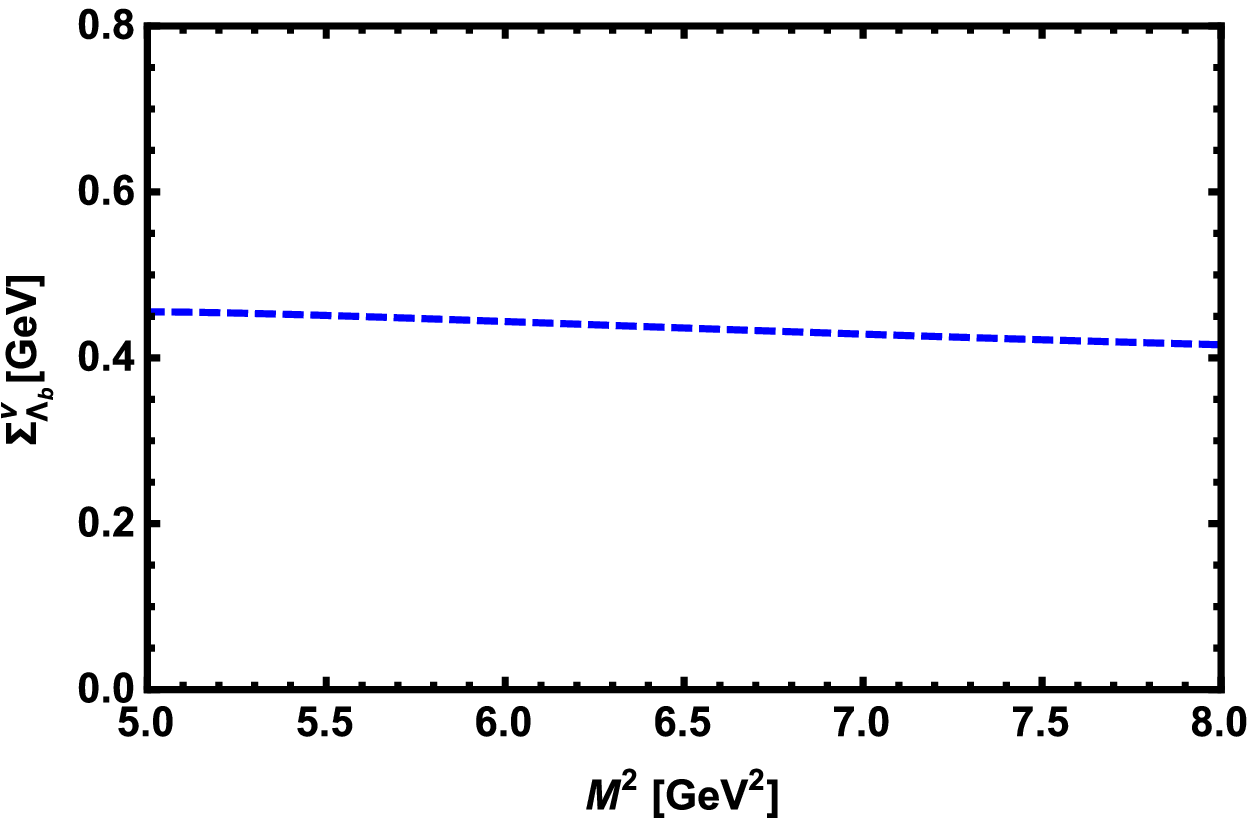,width=0.45\linewidth,clip=} &
\epsfig{file=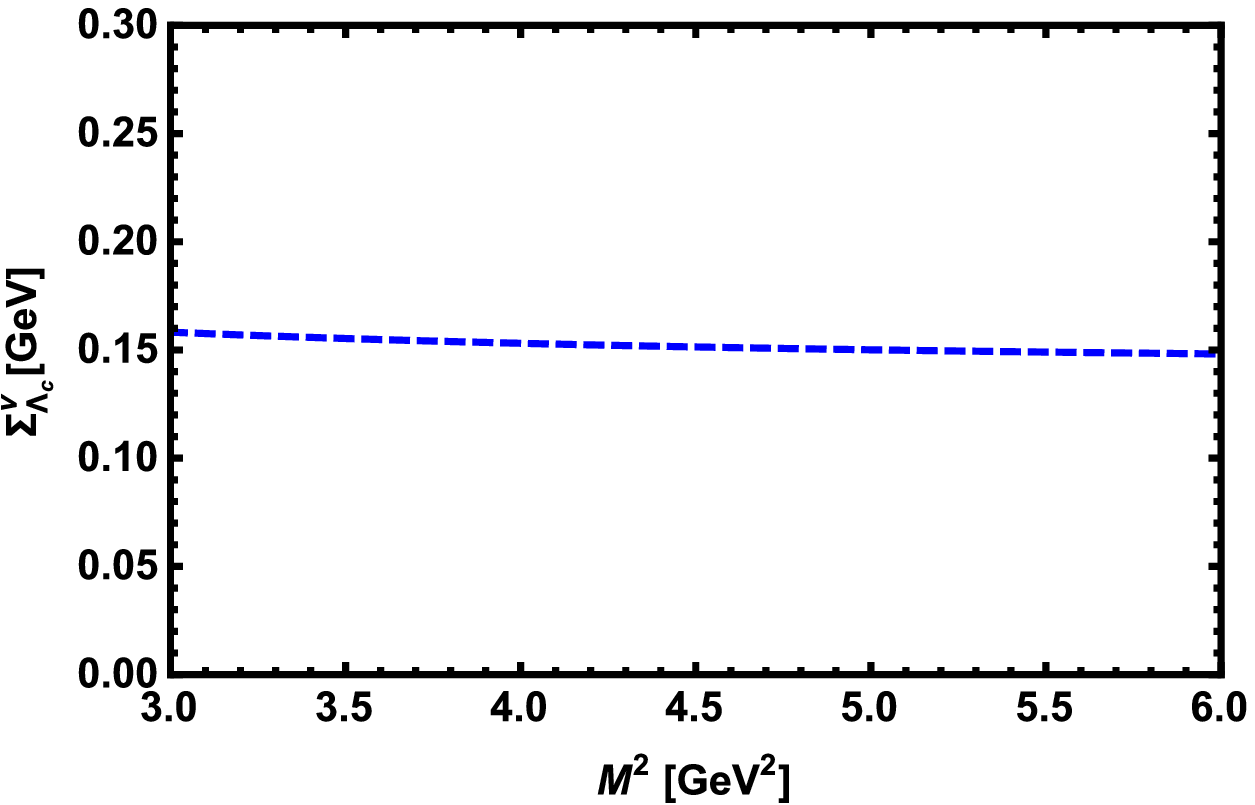,width=0.45\linewidth,clip=} 
\end{tabular}
\caption{The vector self-energy of the $\Lambda_b$  baryon (left panel) and  the $\Lambda_c$  baryon (right panel) versus  $M^2$  in nuclear medium at average values of $s_0$ and $x$.}
\end{figure}
\end{widetext}

\begin{widetext}

\begin{table}[ht!]
\centering
\begin{tabular}{|c|c|c|c|c|c|c|}
\hline \hline
 &$\lambda_{\Lambda_{b}}$ [GeV$^3]$  &$\lambda^{*}_{\Lambda_{b}}$ [GeV$^3]$  & $m_{\Lambda_{b}}$  [GeV] & $m^{*}_{\Lambda_{b}}$  [GeV] & $\Sigma^{\nu}_{ \Lambda_{b}}$ [MeV] & $\Sigma^{S}_{\Lambda_{b}}$ [MeV] \\
 \hline
 Present study & $0.044\pm 0.008$ & $ 0.032\pm 0.006$ & $5.614\pm 0.345$ & $ 6.267 \pm 0.175 $ & $  436\pm 148$ &653  \\  
Ioffe current \cite{Wang:2011hta} &$ 0.027\pm 0.003 $&$ 0.026\pm 0.003 $ &$5.618^{+0.078}_{-0.104}  $  &$ 5.678^{+0.077}_{-0.103}  $ & $  32 \pm 1$ & 60 \\  \hline \hline
&$\lambda_{\Lambda_{c}}$ [GeV$^3]$  &$\lambda^{*}_{\Lambda_{c}}$ [GeV$^3]$  & $m_{\Lambda_{c}}$  [GeV] & $m^{*}_{\Lambda_{c}}$  [GeV] & $\Sigma^{\nu}_{ \Lambda_{c}}$ [MeV] & $\Sigma^{S}_{\Lambda_{c}}$ [MeV] \\
 \hline
Present study  & $0.044\pm 0.012$&$ 0.035\pm 0.010$ & $ 2.295 \pm 0.251 $ & $ 2.567 \pm 0.362 $ & $  151\pm 45 $ & 272  \\
Ioffe current \cite{Wang:2011hta} &$ 0.022\pm 0.002 $&$ 0.021\pm 0.001 $ & $ 2.284^{+0.049}_{-0.078}  $& $ 2.335^{+0.045}_{-0.072}  $ & $  34 \pm 1$ & 51 \\  
 \hline \hline
\end{tabular}
\caption{The numerical values of residues, masses in vacuum and nuclear matter, vector self-energies in nuclear matter  and scalar self-energies of  $\Lambda_b$ and $\Lambda_c$ baryons.}
\end{table}
\begin{table}[ht!]
\centering
\begin{tabular}{|c|c|c|c|c|c|c|}
\hline \hline
 &$\lambda_{\Sigma_{b}}$ [GeV$^3]$  &$\lambda^{*}_{\Sigma_{b}}$ [GeV$^3]$  & $m_{\Sigma_{b}}$  [GeV] & $m^{*}_{\Sigma_{b}}$  [GeV] & $\Sigma^{\nu}_{ \Sigma_{b}}$ [MeV] & $\Sigma^{S}_{\Sigma_{b}}$ [MeV] \\
 \hline
Present study &$ 0.031\pm0.010 $ &$ 0.017\pm 0.005$ &$ 5.810 \pm 0.241 $ & $ 5.196 \pm 0.667 $ & $  382\pm  129$ & -614  \\  
Ioffe current \cite{Wang:2011yj} & $ 8.73^{+0.90}_{-0.65 } \times 10^{-3} $ & $ 1.25^{+0.08}_{-0.04}\times 10^{-2} $ & $ 3.56 ^{+0.14}_{-0.10} $ & $ 3.33^{+0.09}_{-0.07}  $ & $  776 ^{+42}_{-35}$ & -375  \\  \hline
 &$\lambda_{\Sigma_{c}}$ [GeV$^3]$  &$\lambda^{*}_{\Sigma_{c}}$ [GeV$^3]$  & $m_{\Sigma_{c}}$  [GeV] & $m^{*}_{\Sigma_{c}}$  [GeV] & $\Sigma^{\nu}_{ \Sigma_{c}}$ [MeV] & $\Sigma^{S}_{\Sigma_{c}}$ [MeV] \\
 \hline
Present study & $ 0.021\pm0.007 $ & $ 0.011\pm 0.004$ & $ 2.451 \pm 0.208 $ & $ 1.515 \pm 0.321 $ & $  486\pm  144$ & -936 \\
Ioffe current \cite{Wang:2011yj} &  $ 1.99^{+0.29}_{-0.26}\times 10^{-2} $&  $ 2.46^{+0.22}_{-0.16} \times 10^{-2}$ & $ 1.40^{+0.08}_{-0.05}  $& $ 1.33^{+0.06}_{-0.03}  $ & $  446 ^{+35}_{-27}$ & -123 \\  \hline \hline
\end{tabular}
\caption{The numerical values of residues, masses in vacuum and nuclear matter, vector self-energies in nuclear matter  and scalar self-energies of  $\Sigma_b$ and $\Sigma_c$ baryons.}
\end{table}
\begin{table}[ht!]
\centering
\begin{tabular}{|c|c|c|c|c|c|c|}
\hline \hline
 &$\lambda_{\Xi_{b}}$ [GeV$^3]$  &$\lambda^{*}_{\Xi_{b}}$ [GeV$^3]$  & $m_{\Xi_{b}}$  [GeV] & $m^{*}_{\Xi_{b}}$  [GeV] & $\Sigma^{\nu}_{ \Xi_{b}}$ [MeV] & $\Sigma^{S}_{\Xi_{b}}$ [MeV] \\
 \hline
Present study & $0.054 \pm 0.012 $ & $ 0.050\pm 0.018$ & $ 5.812 \pm 0.179 $ & $ 5.795 \pm 0.127  $ & $  15 \pm 5 $ & -17 \\  \hline
&$\lambda_{\Xi_{c}}$ [GeV$^3]$  &$\lambda^{*}_{\Xi_{c}}$ [GeV$^3]$  & $m_{\Xi_{c}}$  [GeV] & $m^{*}_{\Xi_{c}}$  [GeV] & $\Sigma^{\nu}_{ \Xi_{c}}$ [MeV] & $\Sigma^{S}_{\Xi_{c}}$ [MeV] \\
 \hline
Present study & $ 0.056\pm 0.021 $ & $ 0.054\pm 0.020$ & $ 2.466 \pm 0.118 $ & $ 2.461 \pm 0.130 $& $  1.391\pm 0.529 $ & -5  \\
 \hline \hline
\end{tabular}
\caption{The numerical values of residues, masses in vacuum and nuclear matter, vector self-energies in nuclear matter  and scalar self-energies of  $\Xi_b$ and $\Xi_c$ baryons.}
\end{table}
\end{widetext}

A quick glance at these tables leads to the following results:
\begin{itemize}
  \item the values of masses obtained in vacuum  ($m_{B_Q}$) are in nice consistencies with the experimental data presented in table II and borrowed from PDG \cite{PDG}.
The results of vacuum residues $\lambda_{B_Q}$ are also consistent with the predictions of, for instance, Ref. \cite{Aliev:2009jt} obtained by employing the vacuum sum rules.

  \item The values for the masses and residues obtained in nuclear medium ($m^{*}_{B_Q}$ and $\lambda^{*}_{B_Q}$) show considerably large  shifts from their vacuum values. 
These shifts are large in $\Lambda_Q$ and $\Sigma_Q$  channels compared to the $\Xi_Q$ baryon. This is an expected situation since $\Lambda_Q$ and $\Sigma_Q$ baryons contain both the up and 
down quarks interacting with nuclear matter of the same quark content, while the $\Xi_Q$ baryon includes only one of them and the strange quark.

  \item The maximum and minimum shifts in the masses  due to nuclear medium belong to $\Sigma_c$ and $\Xi_c$ baryons, respectively.
 The maximum  and minimum deviations of the in-medium residues from their vacuum values belong to $\Sigma_b$ and $\Xi_c$ baryons, respectively.
  \item The maximum vector self-energy belongs to the $\Sigma_c$ baryon, while its minimum value corresponds to $\Xi_c$ channel.
  \item The mass shifts in $\Lambda_{b,c}$  channels are positive against the negative mass shifts in  
  $\Sigma_{b,c}$ and $\Xi_{b,c}$ channels. The sign of shifts in the residues for all channels are negative.
  \item Although our results for the vacuum masses in $\Lambda_{Q}$  channel are consistent with those of  \cite{Wang:2011hta}, our predictions for the vacuum masses $m_{\Sigma_b}$
and $m_{\Sigma_c}$ differ considerably with those of  \cite{Wang:2011yj}. The $\Sigma_{Q}$ baryon masses obtained in  \cite{Wang:2011yj} in the $\rho_N\rightarrow0$ limit are much more smaller 
than the experimental data due to the unequal pole residues from different structures considered in this work, therefore the authors normalize the
masses  to the experimental data to obtain the scalar self-energies presented in table IV. 
\item The in-medium results for masses, residues and self energies as well as the vacuum residues overall show large discrepancies between our predictions and those
 of \cite{Wang:2011hta, Wang:2011yj}. This can be attributed to the different working regions of auxiliary parameters used in \cite{Wang:2011hta, Wang:2011yj}. Remind that the Ioffe current used in these works remains 
outside of the reliable region for the  mixing parameter entered to the  currents in our case. When comparing the sign of the self energies in our
 work and  the existing predictions from \cite{Wang:2011hta, Wang:2011yj}  we see that the vector and scalar self energies have the same signs in these studies.
  \item  The amount of shifts due to nuclear medium in our case are overall large compared to the results of \cite{Wang:2011hta, Wang:2011yj} in   $\Lambda_Q$ and $\Sigma_Q$ channels. 
In \cite{Wang:2011hta} and $\Lambda_Q$ channel some parameters like residues have not even been affected by the medium noticeably.
\item  Comparison of the results with those of $ Q=s $ (hyperons) analyzed in Ref. \cite{Azizi:2015ica} reveals that the shifts in the scalar and vector self energies in  $\Sigma_Q$  and $\Lambda_Q$ channels are grater than those of the hyperons. In $ \Xi $ channel, however, we see an inverse situation.

  
\end{itemize}

\begin{widetext}

\begin{figure}[h!]
\label{fig1}
\centering
\begin{tabular}{ccc}
\epsfig{file=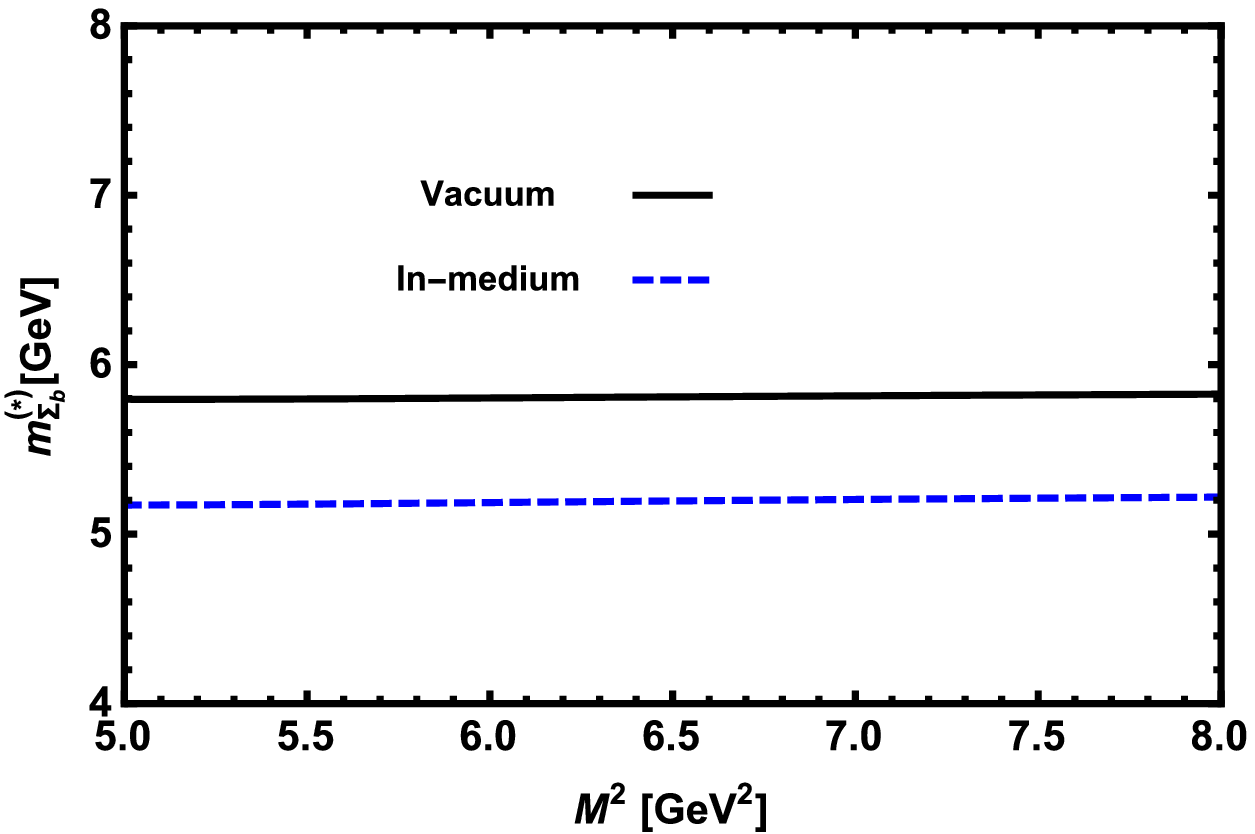,width=0.45\linewidth,clip=} &
\epsfig{file=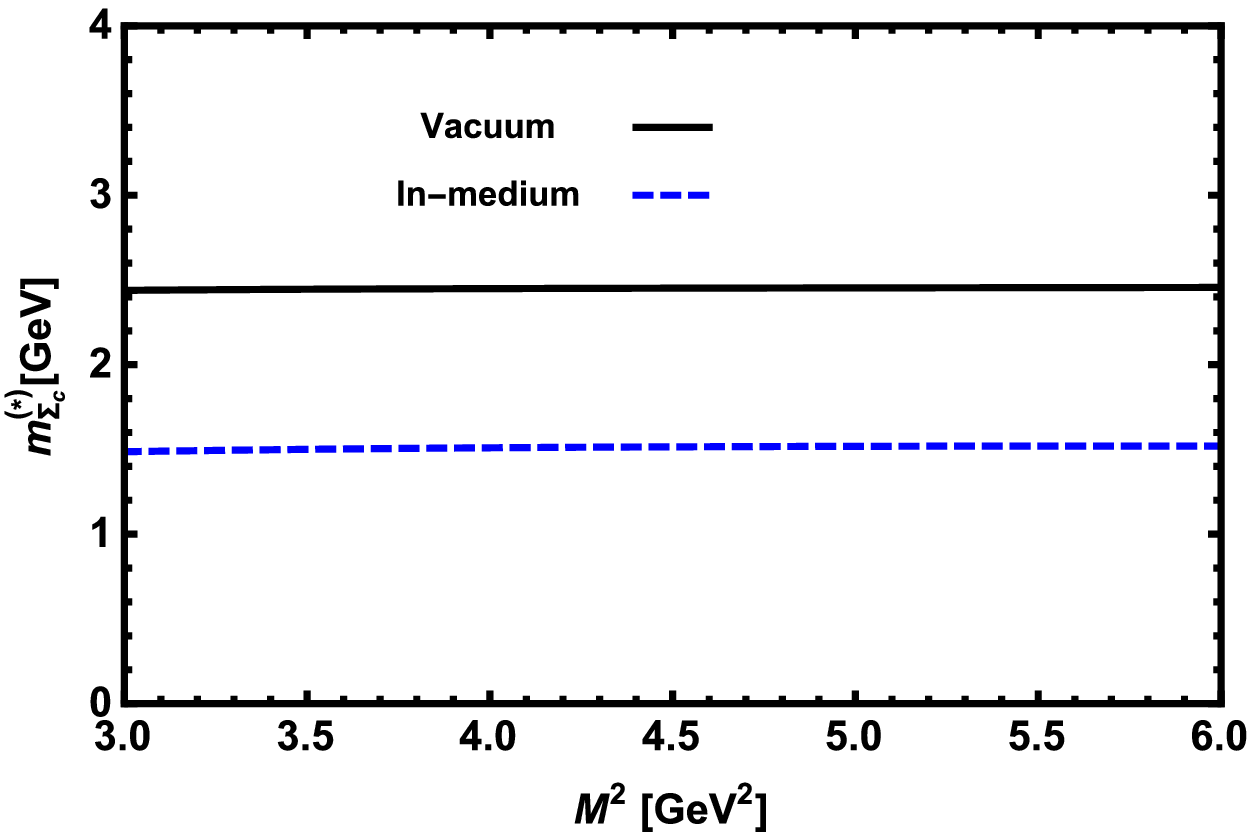,width=0.45\linewidth,clip=} 
\end{tabular}
\caption{The mass of the $\Sigma_b$  baryon (left panel) and  the $\Sigma_c$  baryon (right panel) versus  $M^2$  in vacuum and nuclear medium at average values of $s_0$ and $x$.}
\end{figure}

\begin{figure}[h!]
\label{fig1}
\centering
\begin{tabular}{ccc}
\epsfig{file=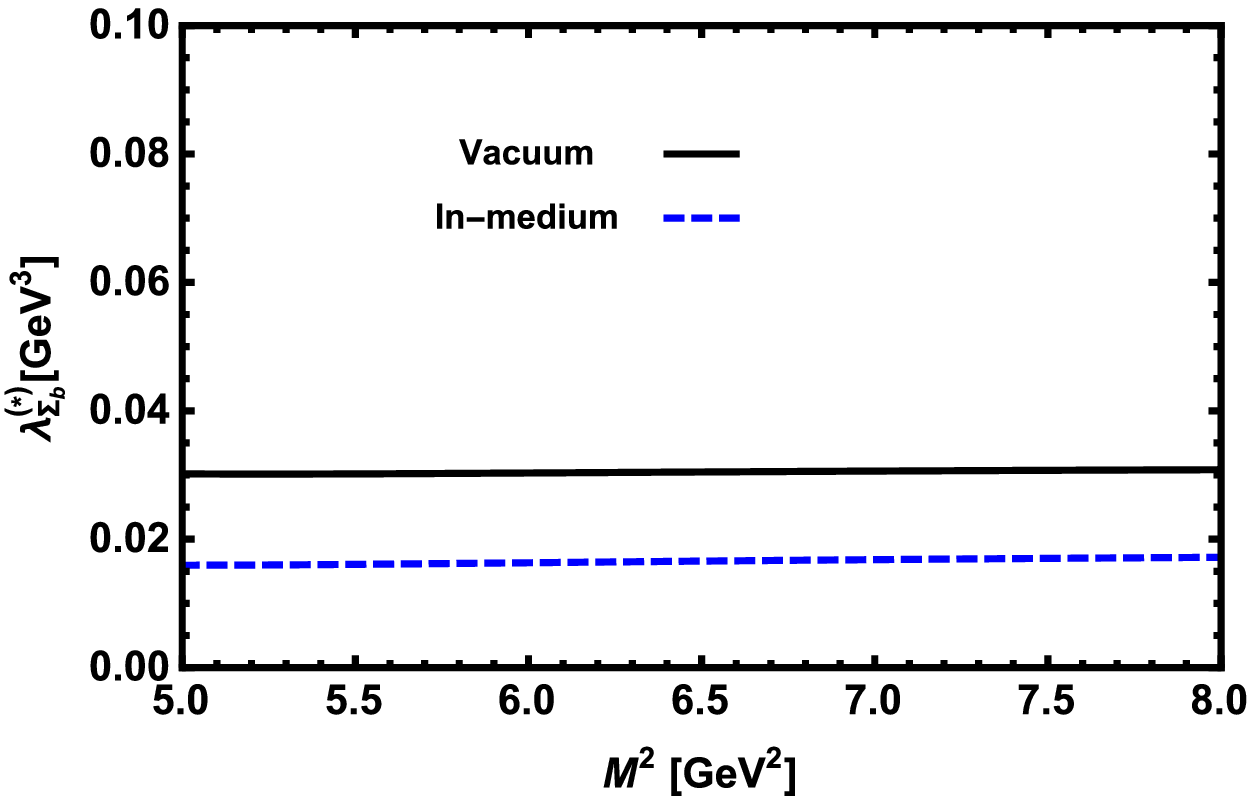,width=0.45\linewidth,clip=} &
\epsfig{file=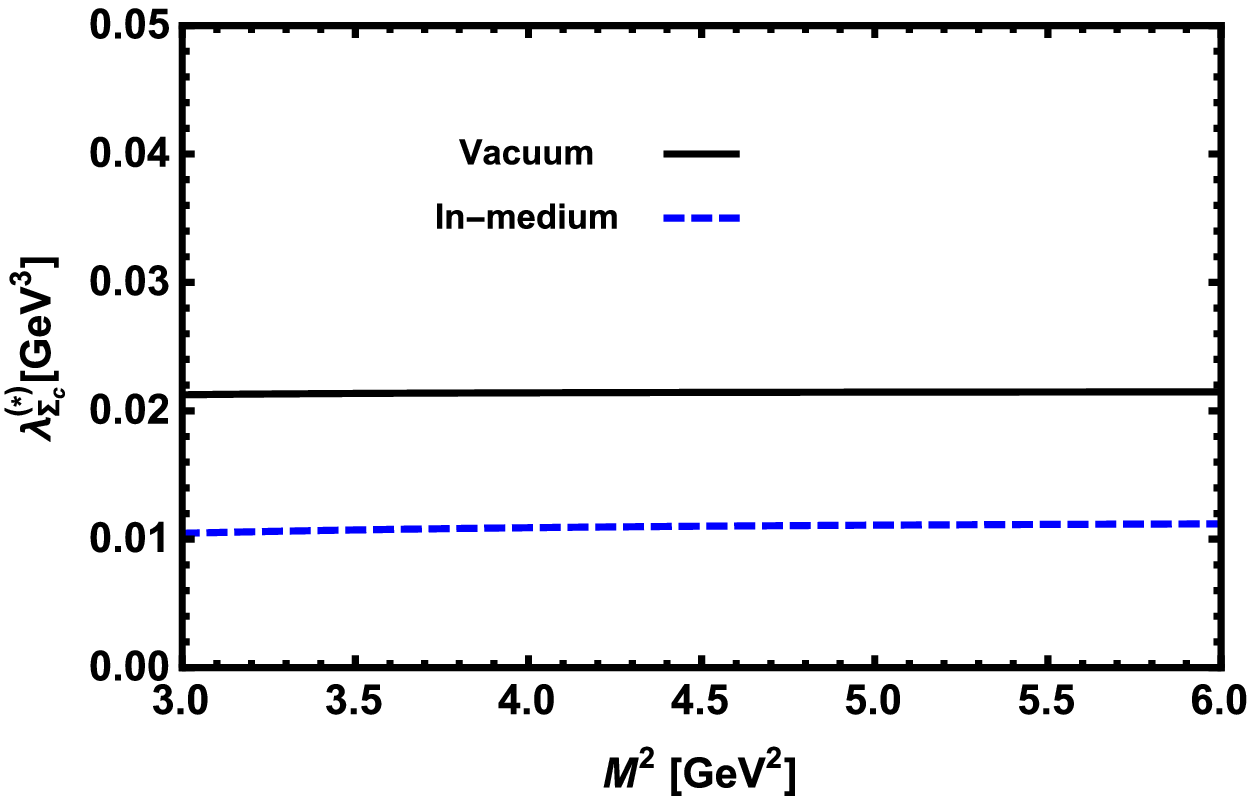,width=0.45\linewidth,clip=} 
\end{tabular}
\caption{The residue of the $\Sigma_b$  baryon (left panel) and  the $\Sigma_c$  baryon (right panel) versus  $M^2$  in vacuum and nuclear medium at average values of $s_0$ and $x$.}
\end{figure}

\begin{figure}[h!]
\label{fig1}
\centering
\begin{tabular}{ccc}
\epsfig{file=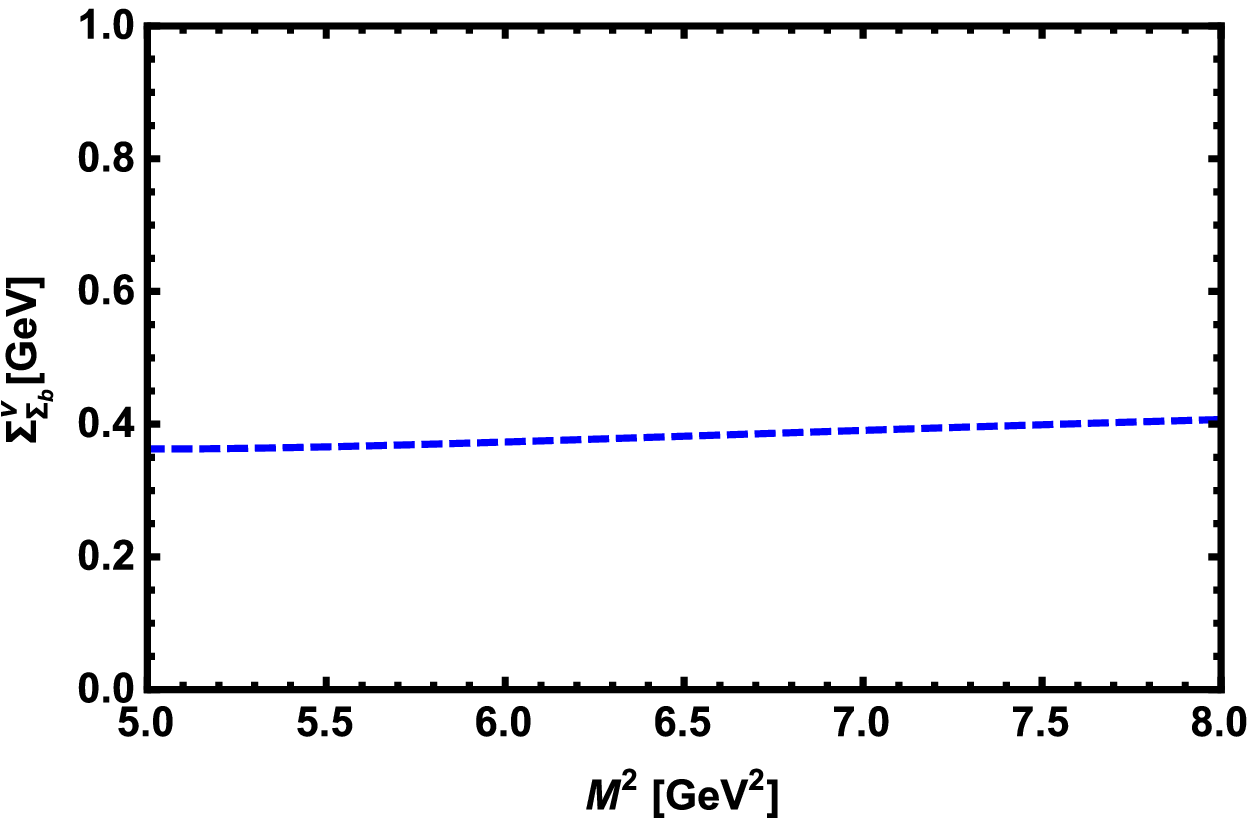,width=0.45\linewidth,clip=} &
\epsfig{file=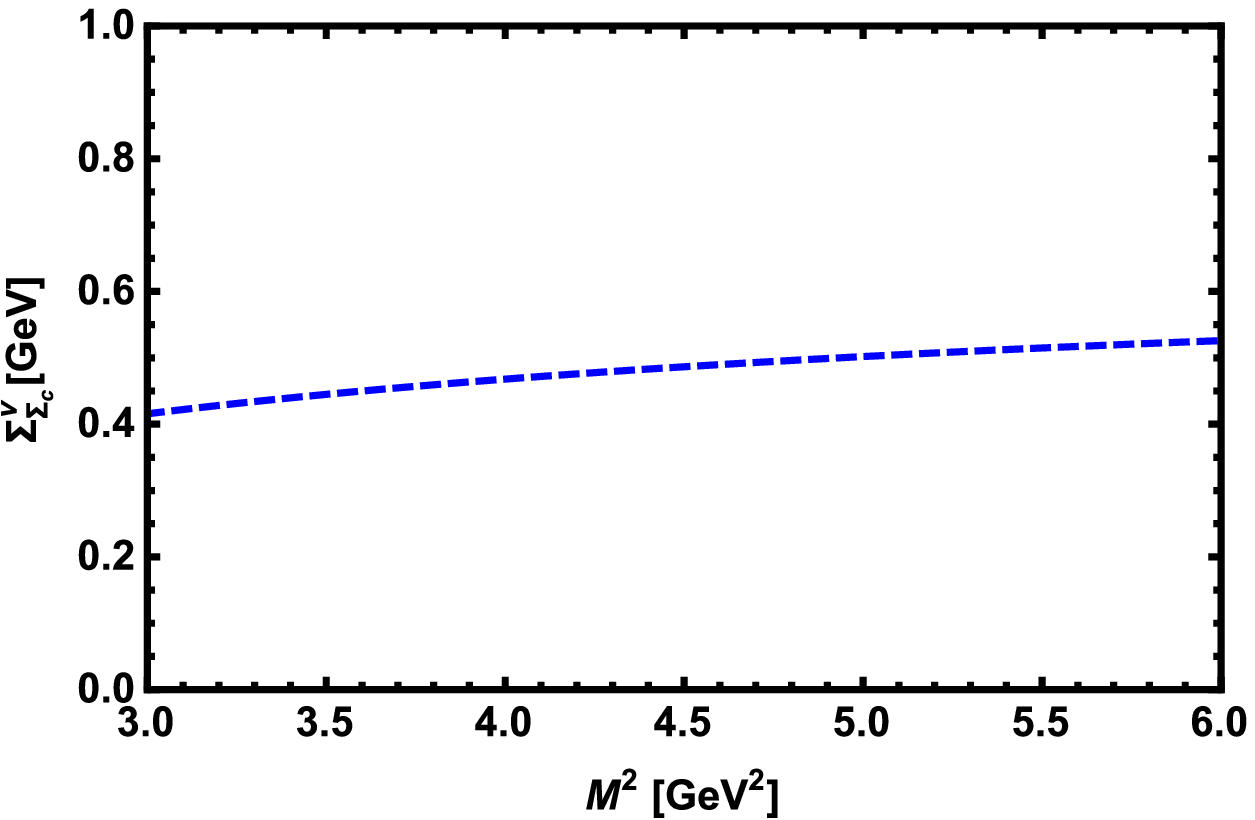,width=0.45\linewidth,clip=} 
\end{tabular}
\caption{The vector self-energy of the $\Sigma_b$  baryon (left panel) and  the $\Sigma_c$  baryon (right panel) versus  $M^2$  in  nuclear medium at average values of $s_0$ and $x$.}
\end{figure}
\end{widetext}

\begin{widetext}

\begin{figure}[h!]
\label{fig1}
\centering
\begin{tabular}{ccc}
\epsfig{file=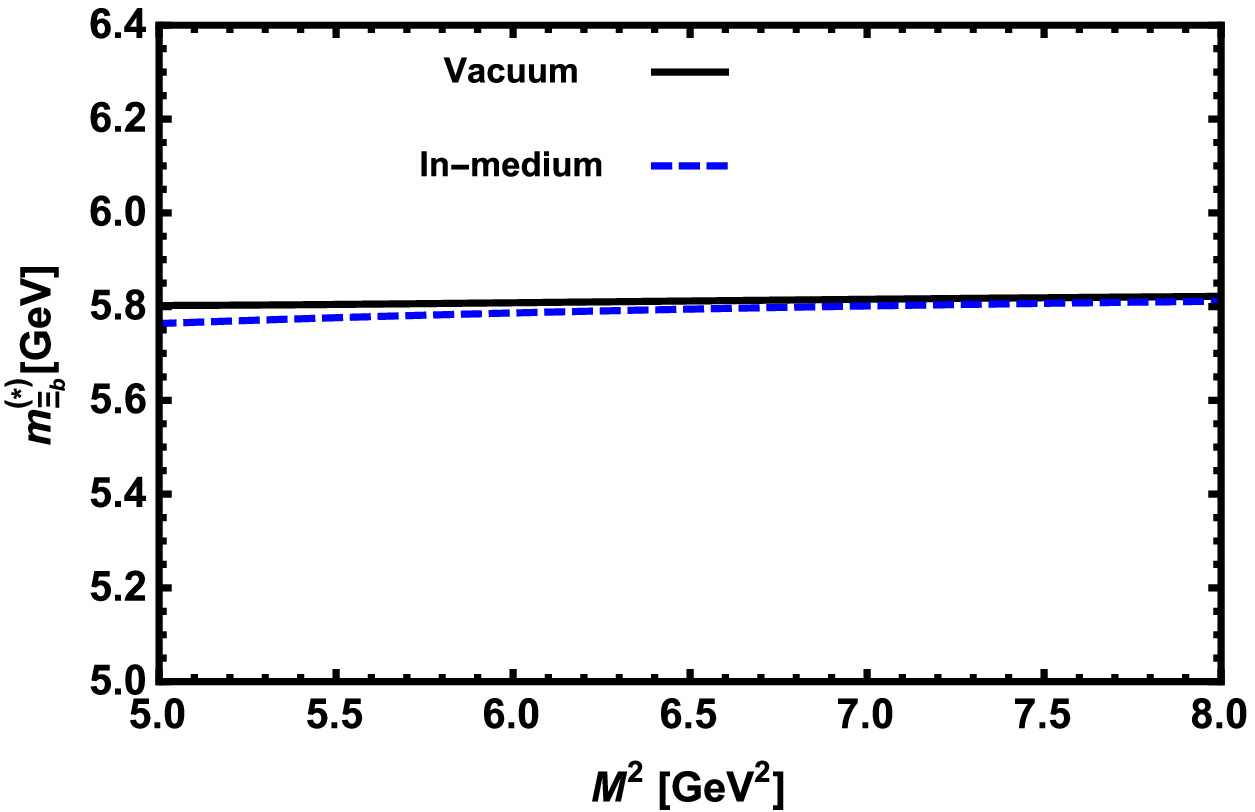,width=0.45\linewidth,clip=} &
\epsfig{file=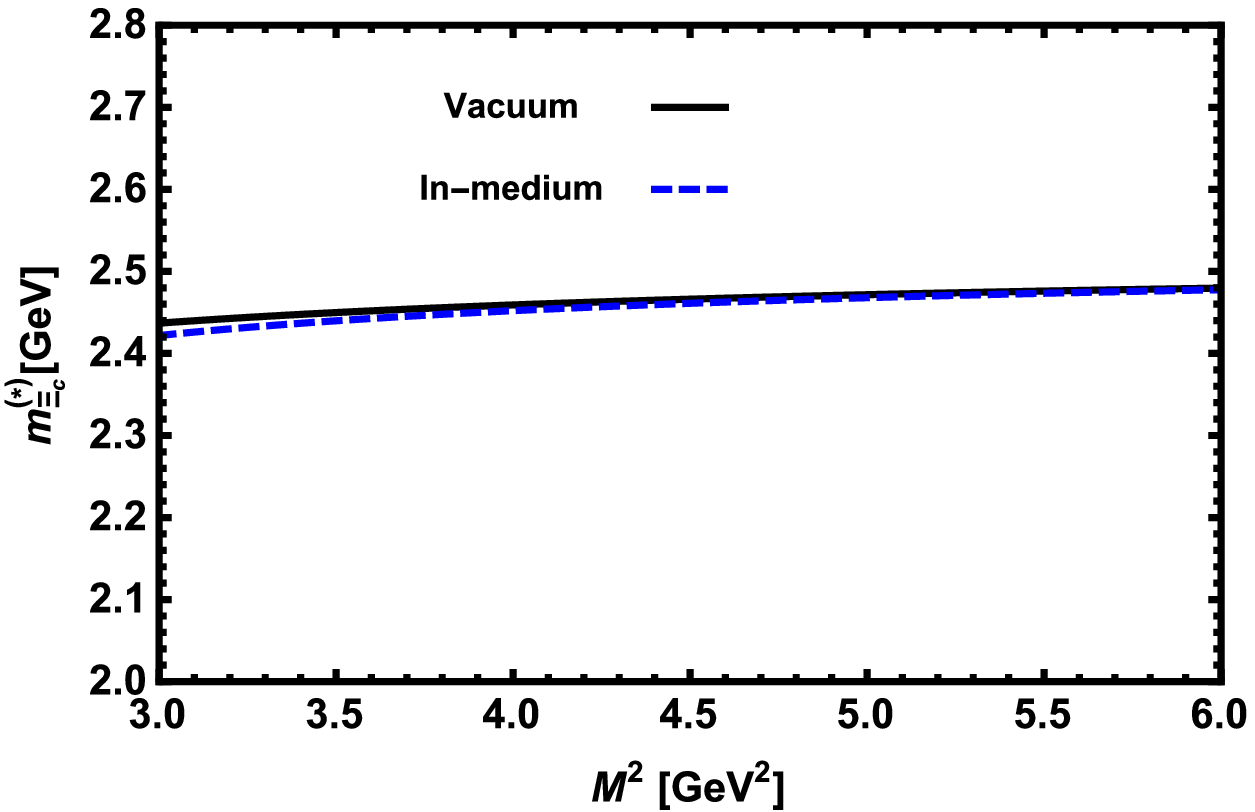,width=0.45\linewidth,clip=} 
\end{tabular}
\caption{The mass of the $\Xi_b$  baryon (left panel) and  the $\Xi_c$  baryon (right panel) versus  $M^2$  in vacuum and nuclear medium at average values of $s_0$ and $x$.}
\end{figure}

\begin{figure}[h!]
\label{fig1}
\centering
\begin{tabular}{ccc}
\epsfig{file=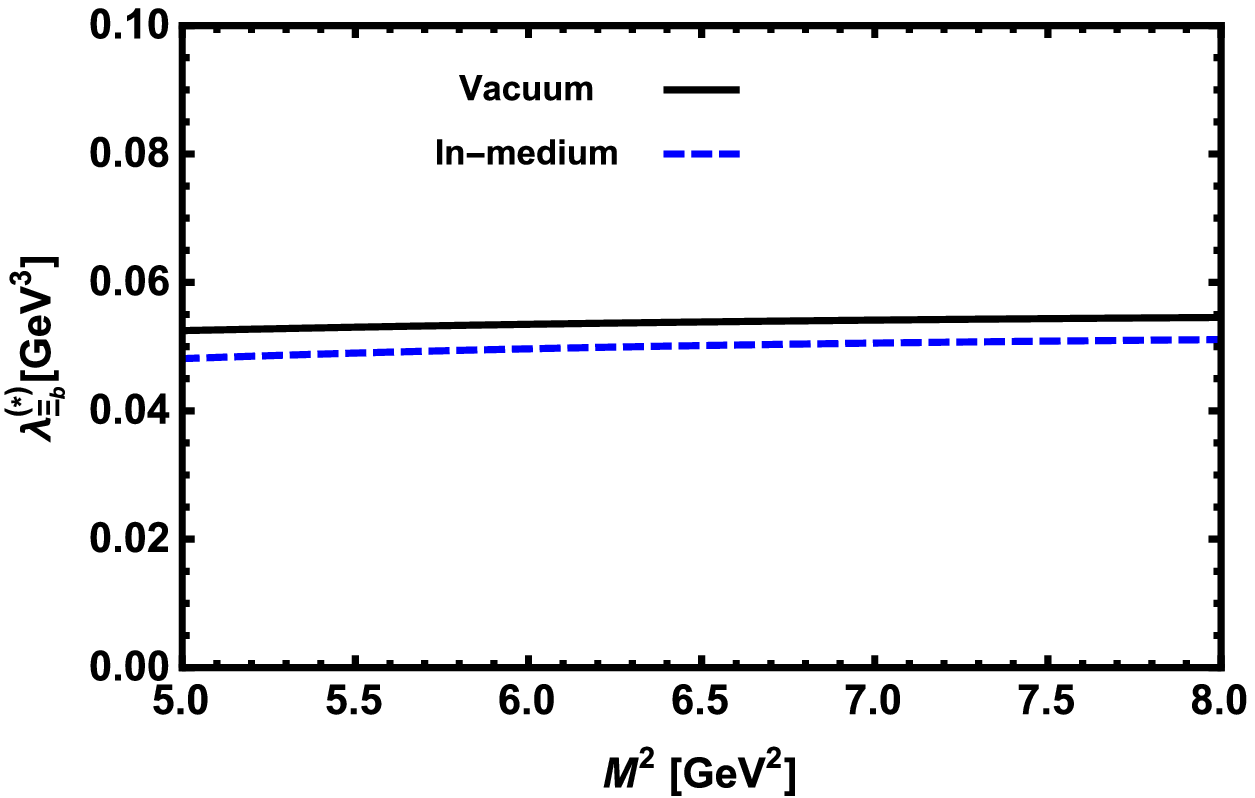,width=0.45\linewidth,clip=} &
\epsfig{file=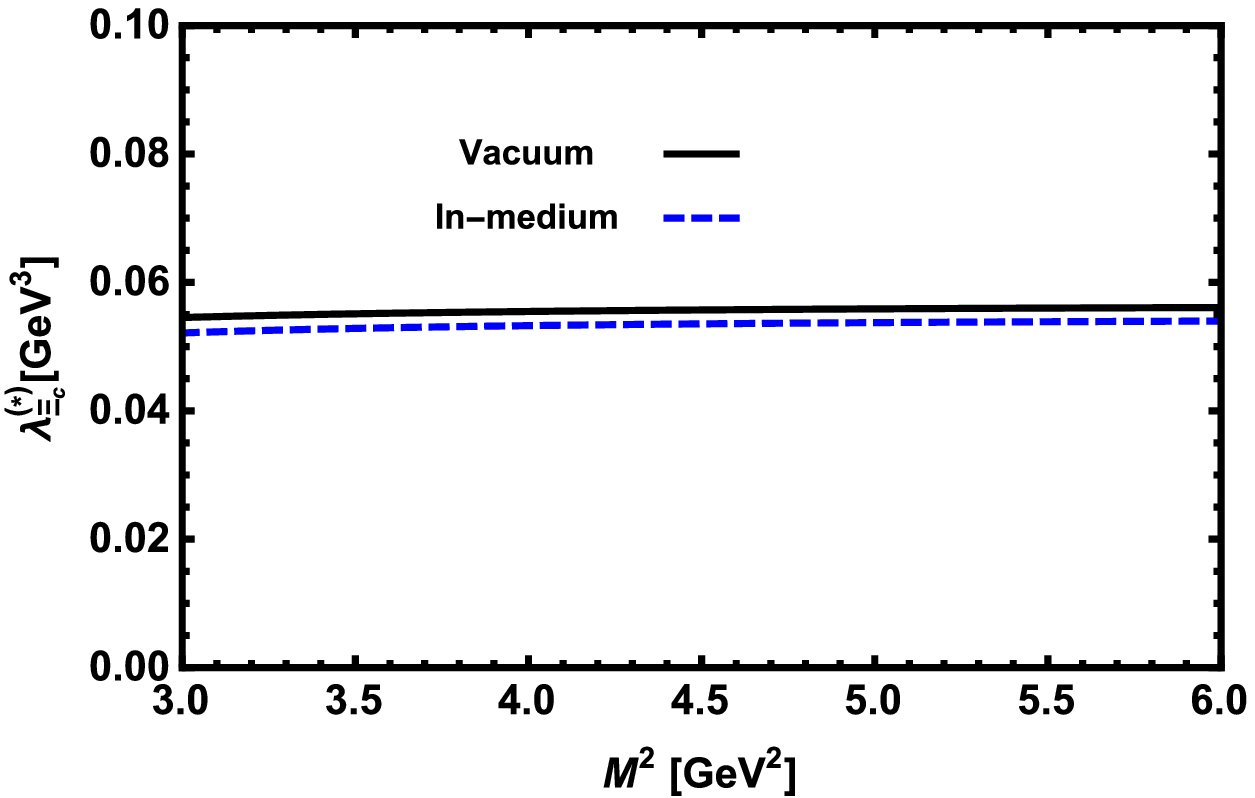,width=0.45\linewidth,clip=} 
\end{tabular}
\caption{The residue of the $\Xi_b$  baryon (left panel) and  the $\Xi_c$  baryon (right panel) versus  $M^2$  in vacuum and nuclear medium at average values of $s_0$ and $x$.}
\end{figure}

\begin{figure}[h!]
\label{fig1}
\centering
\begin{tabular}{ccc}
\epsfig{file=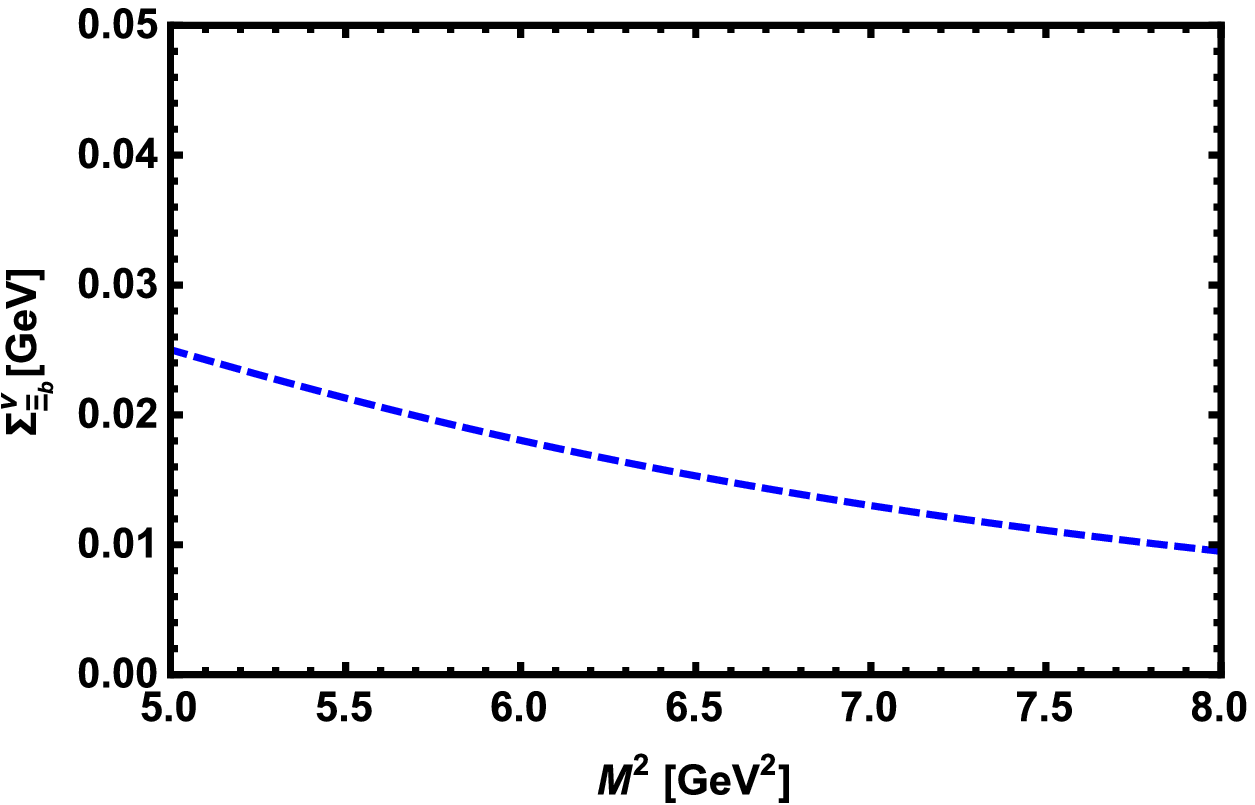,width=0.45\linewidth,clip=} &
\epsfig{file=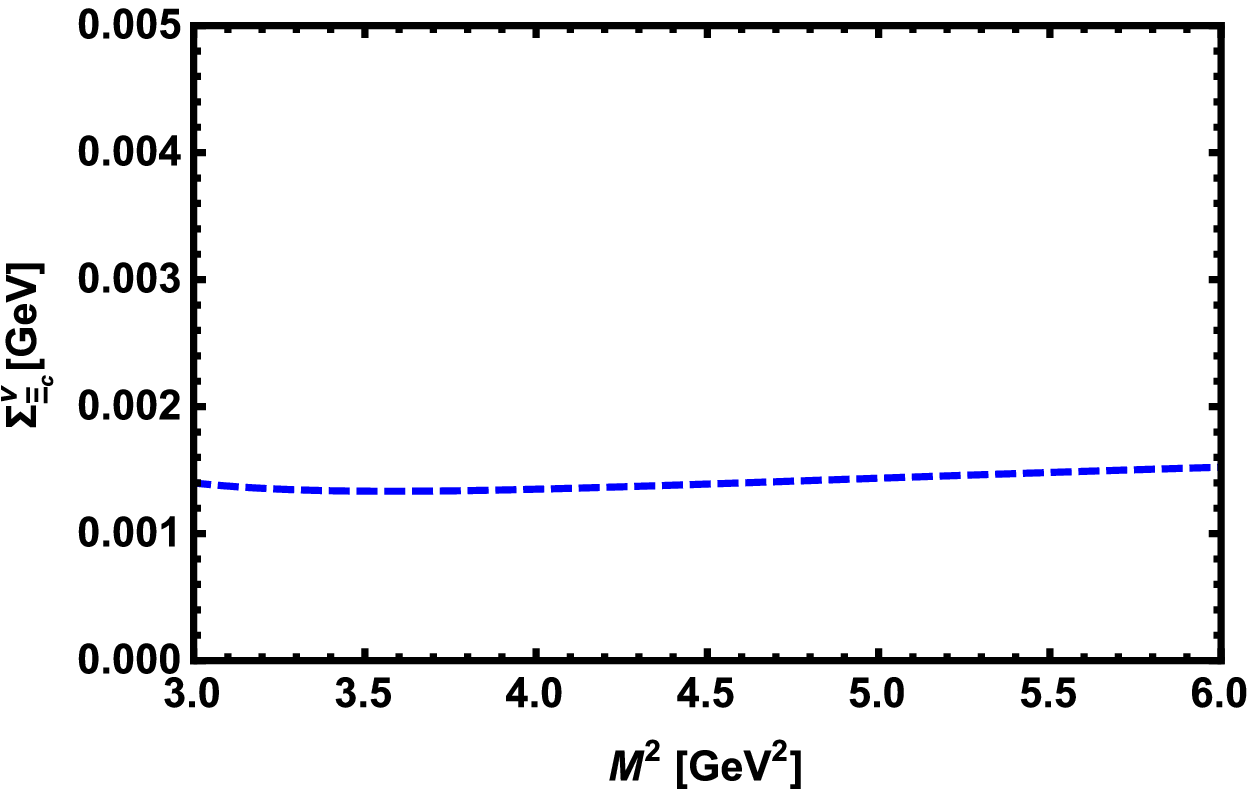,width=0.45\linewidth,clip=} 
\end{tabular}
\caption{The vector self-energy of the $\Xi_b$  baryon (left panel) and  the $\Xi_c$  baryon (right panel) versus  $M^2$  in  nuclear medium at average values of $s_0$ and $x$.}
\end{figure}
\end{widetext}

At the end of this section,  we would like to discuss the density dependence of the results. Note that in above analyses we have used the linear density approximation in operators listed in table II  and the saturation density $ \rho^{sat}_{N} =(0.11)^3  ~GeV^3$   to obtain the numerical results. As an example, we plot $ \lambda^*_{\Lambda_b}/\lambda_{\Lambda_b}$  versus $\rho_N/\rho^{sat}_N$ at the average values of auxiliary parameters in figure 11. $ \lambda^*_{\Lambda_b}/\lambda_{\Lambda_b}$  better represents  the  density-dependence of the OPE for the structure $\!\not\!{p}$ (see Eq. (18)) normalized by a constant, i.e. $ \lambda_{\Lambda_b} $ (residue in vacuum) as  it is proportional to the function $ \widehat{\textbf{B}}\Pi^{QCD}_{p} $ not any ratio of two OPE expressions. From this figure we see that the dependence of the quantity $ \lambda^*_{\Lambda_b}/\lambda_{\Lambda_b}$  on density is roughly linear and it decreases with the increasing the density, considerably.

\begin{figure}[h!]
\label{fig1}
\centering
\begin{tabular}{ccc}
\epsfig{file=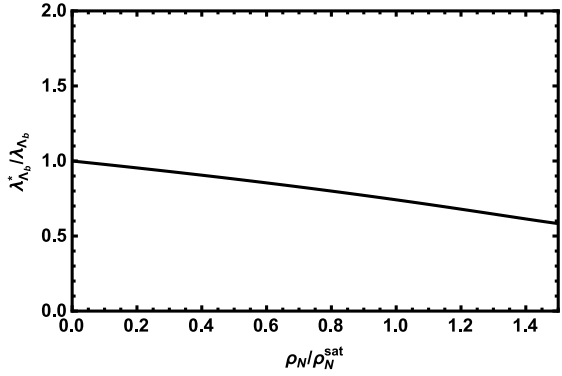,width=0.85\linewidth,clip=}  
\end{tabular}
\caption{$ \lambda^*_{\Lambda_b}/\lambda_{\Lambda_b}$  versus $\rho_N/\rho^{sat}_N$ at the average values of auxiliary parameters. }
\end{figure}

\section{Conclusion}
In the present study, using the interpolating currents with an arbitrary mixing parameter, the in-medium sum rules are utilized to calculate the shifts in the mass and residue as well as the scalar 
and vector self-energies of the heavy $\Lambda_Q, \Sigma_Q$ and $\Xi_Q$ baryons. Imposing the conditions of OPE convergence and pole dominance, we found working regions of  the auxiliary parameters entered the calculations. The obtained results reveal that the shifts in the masses of $\Lambda_{b,c}$  are found to be positive, while  the shits  in the residues of these baryons as well as the shifts in the masses and residues of the $\Sigma_{b,c}$  and $\Xi_{b,c}$ baryons are obtained to be negative.  Both the $ b $ and $ c $ baryons are considerably affected by the medium. 
The maximum shift in mass due to nuclear medium corresponds 
to the  $\Sigma_c$ baryon with the value $\Delta m_{\Sigma_{c}}=-936 ~ MeV$. In the case of residue, it is obtained that the residue 
 of  $\Sigma_b$ baryon is maximally affected by the nuclear medium with the shift $\Delta \lambda_{\Sigma_b} = -0.014 ~ GeV^3 $.
 The scalar and vector self-energies are obtained to be $\Sigma^{S}_{\Lambda_b} = 653 ~ MeV$, $\Sigma^{S}_{\Sigma_b} = -614 ~ MeV $,
 $\Sigma^{S}_{\Xi_b} = -17 ~ MeV $, $\Sigma^{S}_{\Lambda_c} = 272  ~ MeV $, $\Sigma^{S}_{\Sigma_c} = -936 ~ MeV $, $\Sigma^{S}_{\Xi_c} = -5 ~ MeV $
and $\Sigma^{\nu}_{\Lambda_b} = 436 \pm 148  ~ MeV $,  $\Sigma^{\nu}_{\Sigma_b} = 382 \pm 129 ~MeV $, $\Sigma^{\nu}_{\Xi_b} =15 \pm 5 ~ MeV$,  $\Sigma^{\nu}_{\Lambda_c} = 151 \pm 45 ~ MeV $, 
  $\Sigma^{\nu}_{\Sigma_c} = 486 \pm 144 ~ MeV $ and  $\Sigma^{\nu}_{\Xi_c} = 1.391 \pm 0.529 ~ MeV $.
 Our results  in the limit of $\rho_N \rightarrow0$ are in a good consistency with the experimental data and existing theoretical predictions.
 We observed that, because of their quark contents, the  $\Xi_Q$ baryons are minimally affected by the nuclear matter of up and down quarks  in comparison with the $\Lambda_Q$ and $\Sigma_Q$ baryons. Comparison of the results with those of hyperons analyzed in Ref. \cite{Azizi:2015ica} reveals that the shifts in the scalar and vector self energies of  $\Sigma_Q$  and $\Lambda_Q$ baryons are grater than those of the hyperons against the  $ \Xi $ channel, at which  the situation is inverse. 

We performed our numerical analyses and obtained the results  in tables III-V using the linear density approximation for different operators entered to the numerical calculations (see table II) and the saturation density $ \rho^{sat}_{N} =(0.11)^3  ~GeV^3$. We discussed the dependence of the results, as an example $ \lambda^*_{\Lambda_b}/\lambda_{\Lambda_b}$ for the structure  $\!\not\!{p}$, on the nuclear density and found the dependence is  roughly linear and the quantity  $ \lambda^*_{\Lambda_b}/\lambda_{\Lambda_b}$ decreases by increasing the density, considerably. 

Within the working regions of the auxiliary parameters the OPE nicely converges, the  perturbative part exceeds the nonperturbative one and the main nonperturbative contribution comes from the two-quark condensate. The two-gluon, mixed and   poorly known four-quark condensates have very small impacts on the numerical results in the case of heavy baryons.

Looking at the obtained numerical results in the present study at zero temperature (cold nuclear matter) we conclude that the shifts in the parameters of the  $\Sigma_Q$  and $\Lambda_Q$  for both heavy $  b$ and $ c $ quarks due to nuclear matter are considerably large and deserve experimental investigations by different collaborations. The shift in the central value of the mass
of the  $\Sigma_c$ with $ 38\% $  is maximal. The shift in the residue of  $\Sigma_b$  channel with amount of $ 45\% $ is the highest among the shifts in the central values of residues.  The vector self energies are also obtained to be considerably large and experimentally measurable in these channels. The parameters of $\Xi_Q$ with both heavy quarks, however, do not show considerably large shifts due to nuclear matter.  If we combine these results with the results of   Ref. \cite{Azizi:2015ica}  on the parameters of hyperons, which also show considerably large shifts due to nuclear medium especially in the case of negative parity hyperons, we can say that the shifts in the parameters of both heavy and light spin-1/2 baryons are overall  remarkably large and experimentally measurable. These results may shed light to the   PANDA and CBM experiments, which  envisage to study  the hyperons and charmed baryons in $ T=0 $ nuclear matter.  The results of the present study can be checked via different phenomenological approaches as well. To compare the results with those of  the heavy ion collision experiments, one need to add the effects of finite temperature to the results obtained in the present study.



\section{ACKNOWLEDGEMENTS}
This work has been supported in part by Abant \. Izzet Baysal University   the Scientific Research Project Unit under the grant no. 2016.03.02.1004. Work of K. A. was also partly financed by Do\v{g}u\c{s} University through the pro ject: BAP 2015-16-D1-B04.


\appendix*
\section{Correlation functions in terms of quark propagators for different heavy baryons}
In this Appendix, we collect the results for correlation functions in QCD side obtained after insertion of the interpolating currents and contracting out the quark fields in OPE side. 
For the function $\Pi^{OPE}_{\Lambda_b}$ we get
\begin{widetext}
\begin{eqnarray}\label{lambda}
\Pi^{OPE}_{\Lambda_b} &=&
-\frac{i}{6}\epsilon_{abc}\epsilon_{a'b'c'}\int d^4 x e^{ipx} \Big\langle \psi_0\Big|\Big\{\Big(2\gamma_{5}S^{ca'}_{b}(x)S'^{ab'}_{u}(x)S^{bc'}_{d}(x)\gamma_{5} +2\gamma_{5}S^{cb'}_{b}(x)S'^{ba'}_{d}(x)S^{ac'}_{u}(x)\gamma_{5}\nonumber \\
&+&\gamma_{5}S^{ca'}_{d}(x)S'^{bb'}_{b}(x) S^{ac'}_{u}(x)\gamma_{5}+2\gamma_{5}S^{ca'}_{d}(x)S'^{ab'}_{u}(x)S^{bc'}_{b}(x)\gamma_{5}+\gamma_{5}S^{cb'}_{u}(x)S'^{aa'}_{b}(x)S^{bc'}_{d}(x)\gamma_{5}\nonumber \\
&+&2\gamma_{5}S^{cb'}_{u}(x)S'^{ba'}_{d}(x)S^{ac'}_{b}(x)\gamma_{5} -  \gamma_{5}S^{cc'}_{u}(x)\gamma_{5}Tr \Big [S^{ab'}_{b}(x)S'^{ba'}_{d}(x)\Big ] - \gamma_{5}S^{cc'}_{d}(x)\gamma_{5}Tr \Big [S^{ab'}_{u}(x)S'^{ba'}_{b}(x)\Big ]\nonumber \\
&-& 4\gamma_{5}S^{cc'}_{b}(x)\gamma_{5}Tr \Big [S^{ab'}_{u}(x)S'^{ba'}_{d}(x)\Big ] \Big) + \beta\Big(2\gamma_{5}S^{ca'}_{b}(x)\gamma_{5}S'^{ab'}_{u}(x)S^{bc'}_{d}(x) +2\gamma_{5}S^{cb'}_{b}(x)\gamma_{5}S'^{ba'}_{d}(x)S^{ac'}_{u}(x) \nonumber \\
&+&\gamma_{5}S^{ca'}_{d}(x)\gamma_{5}S'^{bb'}_{b}(x)S^{ac'}_{u}(x) +2\gamma_{5}S^{ca'}_{d}(x)\gamma_{5}S'^{ab'}_{u}(x)S^{bc'}_{b}(x) +\gamma_{5}S^{cb'}_{u}(x)\gamma_{5}S'^{aa'}_{b}(x)S^{bc'}_{d}(x) \nonumber \\
&+&2\gamma_{5}S^{cb'}_{u}(x)\gamma_{5}S'^{ba'}_{d}(x)S^{ac'}_{b}(x) +2S^{ca'}_{b}(x)S'^{ab'}_{u}(x)\gamma_{5}S^{bc'}_{d}(x)\gamma_{5} +2S^{cb'}_{b}(x)S'^{ba'}_{d}(x)\gamma_{5}S^{ac'}_{u}(x)\gamma_{5} \nonumber \\
&+&S^{ca'}_{d}(x)S'^{bb'}_{b}(x)\gamma_{5}S^{ac'}_{u}(x)\gamma_{5} +2S^{ca'}_{d}(x)S'^{ab'}_{u}(x)\gamma_{5}S^{bc'}_{b}(x)\gamma_{5} + S^{cb'}_{u}(x)S'^{aa'}_{b}(x)\gamma_{5}S^{bc'}_{d}(x)\gamma_{5} \nonumber \\
&+&2S^{cb'}_{u}(x)S'^{ba'}_{d}(x)\gamma_{5}S^{ac'}_{b}(x)\gamma_{5} -\gamma_{5}S^{cc'}_{u}(x)Tr \Big [S^{ab'}_{b}(x)\gamma_5 S'^{ba'}_{d}(x)\Big ]-S^{cc'}_{u}(x)\gamma_5 Tr \Big [S^{ab'}_{b}(x) S'^{ba'}_{d}(x)\gamma_5 \Big ]\nonumber \\
&-&\gamma_{5}S^{cc'}_{d}(x)Tr \Big [S^{ab'}_{u}(x)\gamma_5 S'^{ba'}_{b}(x)\Big ]-4\gamma_{5}S^{cc'}_{b}(x)Tr \Big [S^{ab'}_{u}(x)\gamma_5 S'^{ba'}_{d}(x)\Big ]-S^{cc'}_{d}(x)\gamma_5 Tr \Big [S^{ab'}_{u}(x)S'^{ba'}_{b}(x)\gamma_5 \Big ]\nonumber \\
&-&4S^{cc'}_{b}(x)\gamma_5 Tr \Big [S^{ab'}_{u}(x)S'^{ba'}_{d}(x)\gamma_5 \Big ]\Big) - \beta^2\Big(2S^{ca'}_{b}(x)\gamma_{5}S'^{ab'}_{u}(x)\gamma_5  S^{bc'}_{d}(x) +2S^{cb'}_{b}(x)\gamma_5 S'^{ba'}_{d}(x)\gamma_{5}S^{ac'}_{u}(x)  \nonumber \\
&+& S^{ca'}_{d}(x)\gamma_5 S'^{bb'}_{b}(x)\gamma_{5}S^{ac'}_{u}(x) + 2S^{ca'}_{d}(x) \gamma_5 S'^{ab'}_{u}(x) \gamma_{5} S^{bc'}_{b}(x) + S^{cb'}_{u}(x) \gamma_5 S'^{aa'}_{b}(x) \gamma_{5}S^{bc'}_{d}(x) \nonumber \\
&+& 2 S^{cb'}_{u}(x) \gamma_5 S'^{ba'}_{d}(x) \gamma_{5}S ^{ac'}_{b}(x) -S^{cc'}_{d}(x)Tr \Big [S^{ba'}_{b}(x)\gamma_{5}S'^{ab'}_{u}(x)\gamma_5 \Big ] -S^{cc'}_{u}(x)Tr \Big [S^{ba'}_{d}(x)\gamma_{5}S'^{ab'}_{b}(x)\gamma_5  \Big ] \nonumber \\
&-&4S^{cc'}_{b}(x)Tr \Big [S^{ba'}_{d}(x)\gamma_{5}S'^{ab'}_{u}(x)\gamma_5  \Big ] \Big)\Big\}\Big| \psi_0\Big\rangle. \nonumber \\
\end{eqnarray}
\end{widetext}
Just by replacing the d quark with the s quark in the above equation, we get the correlation function  for the $\Xi_b$ baryon. For $\Pi^{OPE}_{\Sigma_b}$ we get
\begin{widetext}
\begin{eqnarray}\label{sigma}
\Pi^{OPE}_{\Sigma_b} &=&
\frac{i}{2}\epsilon_{abc}\epsilon_{a'b'c'}\int d^4 x e^{ipx} \Big\langle \psi_0\Big|\Big\{\Big(\gamma_{5}S^{ca'}_{d}(x)S'^{bb'}_{b}(x)S^{ac'}_{u}(x)\gamma_{5} +\gamma_{5}S^{cb'}_{u}(x)S'^{aa'}_{b}(x)S^{bc'}_{d}(x)\gamma_{5} \nonumber \\
&+&\gamma_{5}S^{cc'}_{u}(x)\gamma_{5} Tr \Big [S^{ab'}_{b}(x)S'^{ba'}_{d}(x)\Big] +\gamma_{5}S^{cc'}_{d}(x)\gamma_{5} Tr \Big [S^{ab'}_{u}(x)S'^{ba'}_{b}(x) \Big]\Big) +\beta \Big(\gamma_{5}S^{ca'}_{d}(x)\gamma_{5} S'^{bb'}_{b}(x)S^{ac'}_{u}(x)\nonumber \\
&+&\gamma_{5}S^{cb'}_{u}(x)\gamma_{5} S'^{aa'}_{b}(x)S^{bc'}_{d}(x)+S^{ca'}_{d}(x) S'^{bb'}_{b}(x) \gamma_{5} S^{ac'}_{u}(x)\gamma_{5} +S^{cb'}_{u}(x) S'^{aa'}_{b}(x) \gamma_{5} S^{bc'}_{d}(x)\gamma_{5} \nonumber \\
&+&\gamma_{5}S^{cc'}_{u}(x)Tr \Big[ S^{ab'}_{b}(x)\gamma_{5}S'^{ba'}_{d}(x) \Big] +S^{cc'}_{u}(x)\gamma_{5}Tr \Big[ S^{ab'}_{b}(x)S'^{ba'}_{d}(x)\gamma_{5} \Big] +\gamma_{5}S^{cc'}_{d}(x)Tr \Big[ S^{ab'}_{u}(x)\gamma_{5}S'^{ba'}_{b}(x) \Big] \nonumber \\
&+&S^{cc'}_{d}(x)\gamma_{5}Tr \Big[ S^{ab'}_{u}(x)S'^{ba'}_{b}(x)\gamma_{5} \Big] +\beta^2 \Big(S^{ca'}_{d}(x)\gamma_{5} S'^{bb'}_{b}(x)\gamma_{5}S^{ac'}_{u}(x)+S^{cb'}_{u}(x)\gamma_{5} S'^{aa'}_{b}(x)\gamma_{5}S^{bc'}_{d}(x)\nonumber \\
&+&S^{cc'}_{d}(x)Tr \Big[ S^{ba'}_{b}(x)\gamma_{5}S'^{ab'}_{u}(x)\gamma_{5} \Big] +S^{cc'}_{u}(x)Tr \Big[ S^{ba'}_{d}(x)\gamma_{5}S'^{ab'}_{b}(x)\gamma_{5} \Big] \Big )
\Big \}\Big| \psi_0\Big\rangle, \nonumber \\
\end{eqnarray}
\end{widetext}
where $S'_i=CS_i^TC$ with $i=u,d,s$ or $b$ quark and the notation $Tr[...]$ is used for the trace of gamma matrices. The correlation functions for c-baryons are obtained by the naive replacement
$b\rightarrow c$.


\begin{thebibliography}{99}

\bibitem{Prencipe:2015cgg} 
  E.~Prencipe, J.~S.~Lange and A.~Blinov,
  ``New Spectroscopy with PANDA at FAIR: X, Y, Z and the F-wave Charmonium States,''
  arXiv:1512.05496 [hep-ex].
  
\bibitem{Biswas:2015paa} 
  S.~Biswas {\it et al.},
  ``Measurement of the spark probability of a GEM detector for the CBM muon chamber (MuCh),''
  Nucl.\ Instrum.\ Meth.\ A {\bf 800}, 93 (2015)
  [arXiv:1504.00001 [physics.ins-det]].
  
 \bibitem{fair} 	http://www.gsi.de/fair/experiments/CBM/index.e.html
 
\bibitem{panda} http://www-panda.gsi.de/auto/phy/home.htm
 
 \bibitem{Friman} B. Friman et al., The CBM Physics Book: Compressed Baryonic Matter in Laboratory Experiments, Springer, Heidelberg (2011).
 
\bibitem{Lutz:2009ff} 
  M.~F.~M.~Lutz {\it et al.} [PANDA Collaboration],
  ``Physics Performance Report for PANDA: Strong Interaction Studies with Antiprotons,''
  arXiv:0903.3905 [hep-ex].
  
\bibitem{Giacosa:2015nwa} 
 F.~Giacosa,
 ``Non-conventional mesons at PANDA,''
 J.\ Phys.\ Conf.\ Ser.\  {\bf 599}, no. 1, 012004 (2015)
 [arXiv:1502.02682 [hep-ph]].
 
\bibitem{Drukarev:1988kd} 
  E.~G.~Drukarev and E.~M.~Levin,
  ``The QCD Sum Rules and Nuclear Matter. 2.,''
  Nucl.\ Phys.\ A {\bf 511}, 679 (1990)
  Erratum: [Nucl.\ Phys.\ A {\bf 516}, 715 (1990)].
  
\bibitem{Hatsuda:1990zj} 
  T.~Hatsuda, H.~Hogaasen and M.~Prakash,
  ``QCD sum rules and the Okamoto-Nolen-Schiffer anomaly,''
  Phys.\ Rev.\ Lett.\  {\bf 66}, 2851 (1991)
  Erratum: [Phys.\ Rev.\ Lett.\  {\bf 69}, 1290 (1992)].
  
\bibitem{Adami:1991js} 
  C.~Adami and G.~E.~Brown,
  ``Isospin breaking in nuclear physics: The Nolen-Schiffer effect,''
  Z.\ Phys.\ A {\bf 340}, 93 (1991).
  
\bibitem{Jin:1993fr} 
  X.~m.~Jin and R.~J.~Furnstahl,
 ``QCD sum rules for Lambda hyperons in nuclear matter,''
  Phys.\ Rev.\ C {\bf 49}, 1190 (1994).
  
\bibitem{Jin:1994bh} 
  X.~M.~Jin and M.~Nielsen,
  ``QCD sum rules for Sigma hyperons in nuclear matter,''
  Phys.\ Rev.\ C {\bf 51}, 347 (1995)
  [hep-ph/9405331].

\bibitem{Suzuki:2015est} 
  K.~Suzuki, P.~Gubler and M.~Oka,
  ``D meson mass increase by restoration of chiral symmetry in nuclear matter,''
  Phys.\ Rev.\ C {\bf 93}, no. 4, 045209 (2016)
  [arXiv:1511.04513 [hep-ph]].
    
\bibitem{Buchheim:2014uda} 
  T.~Buchheim, T.~Hilger and B.~K\"ampfer,
  ``Heavy-quark expansion for $D$ and $B$ mesons in nuclear matter,''
  EPJ Web Conf.\  {\bf 81}, 05007 (2014)
  [arXiv:1410.0143 [nucl-th]].
  
\bibitem{Buchheim:2015yyc} 
  T.~Buchheim, B.~K\"ampfer and T.~Hilger,
  ``Algebraic vacuum limits of QCD condensates from in-medium projections of Lorentz tensors,''
  J.\ Phys.\ G {\bf 43}, no. 5, 055105 (2016)
  [arXiv:1511.06234 [nucl-th]].
  
\bibitem{Azizi:2014yea} 
  K.~Azizi and N.~Er,
  ``Properties of nucleon in nuclear matter: once more,''
  Eur.\ Phys.\ J.\ C {\bf 74}, 2904 (2014)
  [arXiv:1401.1680 [hep-ph]].
  
\bibitem{Azizi:2015ica} 
  K.~Azizi, N.~Er and H.~Sundu,
  ``Positive and negative parity hyperons in nuclear medium,''
  Phys.\ Rev.\ D {\bf 92}, no. 5, 054026 (2015)
  [arXiv:1506.02183 [hep-ph]].
  
\bibitem{Hayashigaki:2000es} 
  A.~Hayashigaki,
``Mass modification of D meson at finite density in QCD sum rule,''
  Phys.\ Lett.\ B {\bf 487}, 96 (2000)
  [nucl-th/0001051].
  
\bibitem{Hilger:2010zf} 
  T.~Hilger and B.~Kampfer,
``In-Medium Modifications of Scalar Charm Mesons in Nuclear Matter,''
  Nucl.\ Phys.\ Proc.\ Suppl.\  {\bf 207-208}, 277 (2010)
  [arXiv:1011.2627 [nucl-th]].
  
\bibitem{Wang:2011mj} 
  Z.~G.~Wang and T.~Huang,
``In-medium mass modifications of the $D_0$ and $B_0$ mesons with the QCD sum rules,''
  Phys.\ Rev.\ C {\bf 84}, 048201 (2011)
  [arXiv:1107.5889 [hep-ph]].
  
\bibitem{Azizi:2014bba} 
  K.~Azizi, N.~Er and H.~Sundu,
``More about the $B$ and $D$ mesons in nuclear matter,''
  Eur.\ Phys.\ J.\ C {\bf 74}, 3021 (2014)
  [arXiv:1405.3058 [hep-ph]].
  
\bibitem{Wang:2011hta}
  Z.~G.~Wang,
  ``Analysis of the $\Lambda_Q$ baryons in the nuclear matter with the QCD sum rules,''
  Eur.\ Phys.\ J.\ C {\bf 71} (2011) 1816
  [arXiv:1108.4251 [hep-ph]].
  
\bibitem{Wang:2011yj}
  Z.~G.~Wang,
 ``Analysis of the $\Sigma_Q$ baryons in the nuclear matter with the QCD sum rules,''
  Phys.\ Rev.\ C {\bf 85} (2012) 045204
  [arXiv:1109.2180 [hep-ph]].
  
\bibitem{Wang:2012xk} 
  Z.~G.~Wang,
  ``Analysis of the doubly heavy baryons in the nuclear matter with the QCD sum rules,''
  Eur.\ Phys.\ J.\ C {\bf 72}, 2099 (2012)
  [arXiv:1205.0605 [hep-ph]].
  
\bibitem{Thomas:2007gx} 
  R.~Thomas, T.~Hilger and B.~Kampfer,
  ``Four-quark condensates in nucleon QCD sum rules,''
  Nucl.\ Phys.\ A {\bf 795}, 19 (2007)
  [arXiv:0704.3004 [hep-ph]].


\bibitem{Drukarev:2013kga} 
  E.~G.~Drukarev, M.~G.~Ryskin and V.~A.~Sadovnikov,
  ``Nucleon QCD sum rules in instanton medium,''
  J.\ Exp.\ Theor.\ Phys.\  {\bf 121}, no. 3, 408 (2015)
  [arXiv:1312.1449 [hep-ph]].

\bibitem{Leinweber:1994nm} 
  D.~B.~Leinweber,
  ``Nucleon properties from unconventional interpolating fields,''
  Phys.\ Rev.\ D {\bf 51}, 6383 (1995)
  [nucl-th/9406001].


\bibitem{Stein:1994zk} 
  E.~Stein, P.~Gornicki, L.~Mankiewicz, A.~Schafer and W.~Greiner,
  ``QCD sum rule calculation of twist - three contributions to polarized nucleon structure functions,''
  Phys.\ Lett.\ B {\bf 343}, 369 (1995)
  [hep-ph/9409212].


  
\bibitem{Cohen:1991js}
  T.~D.~Cohen, R.~J.~Furnstahl and D.~K.~Griegel,
  ``From QCD sum rules to relativistic nuclear physics,''
  Phys.\ Rev.\ Lett.\  {\bf 67} (1991) 961.
  
\bibitem{Cohen:1994wm} 
  T.~D.~Cohen, R.~J.~Furnstahl, D.~K.~Griegel and X.~m.~Jin,
  ``QCD sum rules and applications to nuclear physics,''
  Prog.\ Part.\ Nucl.\ Phys.\  {\bf 35}, 221 (1995)
  [hep-ph/9503315].

\bibitem{Furnstahl:1995nd} 
  R.~J.~Furnstahl, X.~m.~Jin and D.~B.~Leinweber,
  ``New QCD sum rules for nucleons in nuclear matter,''
  Phys.\ Lett.\ B {\bf 387}, 253 (1996)
  [nucl-th/9511007].



\bibitem{PDG} K.A. Olive et al. (Particle Data Group), Chin. Phys. C, 38, 090001 (2014).



\bibitem{Cohen:1991nk} 
  T.~D.~Cohen, R.~J.~Furnstahl and D.~K.~Griegel,
  ``Quark and gluon condensates in nuclear matter,''
  Phys.\ Rev.\ C {\bf 45}, 1881 (1992).
  



\bibitem{Belyaev:1982cd} 
  V.~M.~Belyaev and B.~L.~Ioffe,
  ``Determination of the baryon mass and baryon resonances from the quantum-chromodynamics sum rule. Strange baryons,''
  Sov.\ Phys.\ JETP {\bf 57}, 716 (1983)
  [Zh.\ Eksp.\ Teor.\ Fiz.\  {\bf 84}, 1236 (1983)].
  
\bibitem{Ioffe:2005ym} 
  B.~L.~Ioffe,
  ``QCD at low energies,''
  Prog.\ Part.\ Nucl.\ Phys.\  {\bf 56}, 232 (2006)
  [hep-ph/0502148].
  
  \bibitem{Thomas1} A. W. Thomas, P. E. Shanahan, R. D. Young, Nuovo Cim. C 035N04, 3 (2012).

\bibitem{Dinter:2011za} 
  S.~Dinter, V.~Drach and K.~Jansen,
  ``Dark matter search and the scalar quark contents of the nucleon,''
  Int.\ J.\ Mod.\ Phys.\ Proc.\ Suppl.\ E {\bf 20}, 110 (2011)
  [arXiv:1111.5426 [hep-lat]].
  
  

\bibitem{Aliev:2009jt} 
  T.~M.~Aliev, K.~Azizi and A.~Ozpineci,
  ``Radiative Decays of the Heavy Flavored Baryons in Light Cone QCD Sum Rules,''
  Phys.\ Rev.\ D {\bf 79}, 056005 (2009)
  [arXiv:0901.0076 [hep-ph]].

\end{thebibliography}
\end{document}